\documentclass[10pt,journal,compsoc]{IEEEtran}

\usepackage{url}
\usepackage{graphicx}
\usepackage[caption=false,font=footnotesize]{subfig}
\usepackage[justification=centering]{caption}
\usepackage{multirow}
\usepackage{color}
\usepackage{float}
\usepackage{amsmath}

\begin{document}
\title{SeqTrans: Automatic Vulnerability Fix via Sequence to Sequence Learning}
%
%
%
%

\author{Jianlei~Chi,
        Yu~Qu,
        Ting~Liu,~\IEEEmembership{Member,~IEEE,}
        Qinghua~Zheng,~\IEEEmembership{Member,~IEEE,}
        Heng~Yin,~\IEEEmembership{Member,~IEEE}
\IEEEcompsocitemizethanks{
\IEEEcompsocthanksitem J. Chi, T. Liu and Q. Zheng are with the Ministry of Education Key Lab For Intelligent Networks and Network Security (MOEKLINNS), School of Computer Science and Technology, Xian Jiaotong University, Xian 710049, China.
\protect\\
Email: chijianlei@stu.xjtu.edu.cn, tliu, qhzhengg@xjtu.edu.cn.
\IEEEcompsocthanksitem Y. Qu and H. Yin are with the Department of Computer Science and Engineering, UC Riverside, California, USA.
\protect\\Email: yuq@ucr.edu, heng@cs.ucr.edu
}
}

\markboth{Journal of \LaTeX\ Class Files,~Vol.~14, No.~8, May~2021}%
{Shell \MakeLowercase{\textit{et al.}}: Bare Advanced Demo of IEEEtran.cls for IEEE Computer Society Journals}

\IEEEtitleabstractindextext{
\begin{abstract}
Software vulnerabilities are now reported unprecedentedly due to the recent development of automated vulnerability hunting tools. 
However, fixing vulnerabilities still mainly depends on programmers' manual efforts. Developers need to deeply understand the vulnerability and affect the system's functions as little as possible.

In this paper, with the advancement of Neural Machine Translation (NMT) techniques, we provide a novel approach called SeqTrans to exploit historical vulnerability fixes to provide suggestions and automatically fix the source code. 
To capture the contextual information around the vulnerable code, we propose to leverage data-flow dependencies to construct code sequences and feed them into the state-of-the-art transformer model.
The fine-tuning strategy has been introduced to overcome the small sample size problem.  
We evaluate SeqTrans on a dataset containing 1,282 commits that fix 624 CVEs in 205 Java projects. 
Results show that the accuracy of SeqTrans outperforms the latest techniques and achieves $23.3\%$ in statement-level fix and $25.3\%$ in CVE-level fix.
In the meantime, we look deep inside the result and observe that the NMT model performs very well in certain kinds of vulnerabilities like CWE-287 (Improper Authentication) and CWE-863 (Incorrect Authorization).
\end{abstract}

}

\maketitle

\IEEEraisesectionheading{\section{Introduction}}
\label{sec:introduction}

\IEEEPARstart{S}{oftware} evolves quite frequently for numerous reasons such as deprecating old features, adding new features, refactoring, bug fixing, etc.
Debugging is one of the most time-consuming and painful processes in the entire software development life cycle (SDLC).
A recent study indicates that the debugging component can account for up to 50\% of the overall software development overhead, and the majority of the debugging costs come from manually checking and fixing bugs~\cite{britton2012quantify}. 
This leads to a growing number of researchers working on teaching machines to automatically modify and fix the program, which is called automated program repair~\cite{hailpern2002software, zeller2009programs, gazzola2017automatic, monperrus2018automatic, weimer2009automatically, nguyen2013semfix, kim2013automatic, dallmeier2009generating, marcote2015automatic, ackling2011evolving, long2015staged, qi2014strength, dinella2020hoppity}.

A software vulnerability is one kind of bug that can be exploited by an attacher to cross authorization boundaries in the source code.
Vulnerabilities like HeartBleed~\cite{durumeric2014matter}, Spectre~\cite{kocher2019spectre} and Meltdown~\cite{lipp2018meltdown}, introduced significant threats to millions of users. 
Nevertheless, identifying and fixing vulnerabilities is more challenging than bugs~\cite{potter2004software, scandariato2014predicting, shen2020survey}.
Firstly, the number of vulnerabilities is fewer than bugs, making learning enough knowledge from historical data more difficult.
In other words, we usually have only a relatively small database of vulnerabilities.
Secondly, labeling and identifying vulnerability requires a mindset of the attacker that may not be available to developers~\cite{morrison2015challenges}.
Thirdly, vulnerabilities are reported at an unprecedented speed due to the recent development of automated vulnerability hunting tools like AFL~\cite{zalewski2010american}, AFLGo~\cite{bohme2017directed}, AFLFast~\cite{bohme2017coverage}. 
Nevertheless, fixing vulnerabilities depends heavily on manually generating repair templates and defining repair rules, which are tedious and error-prone~\cite{ma2017vurle}. 
Automatically learning to generate vulnerability fixes is urgently needed and will significantly improve the efficiency of software development and maintenance processes.

There are many works of automated program repair (APR) or called code migration in both industrial and academic domains~\cite{monperrus2018automatic}. 
Some APR studies focus on automatically generating fix templates or called fix patterns~\cite{xu2019meditor, nguyen2010graph, fazzini2019automated, phan2017statistical, lamothe2018a4}.
Some of APR studies focus on mining similar code changes from historical repair records such as CapGen~\cite{wen2018context} and FixMiner~\cite{koyuncu2020fixminer}.
Other approaches utilize static and dynamic analysis with constraining solving to accomplish patch generation~\cite{nguyen2013semfix, liu2019avatar}.
IDEs also provide specific kinds of automatic changes~\cite{eclipse2009eclipse}. 
For example, refactoring, generating getters and setters, adding override/implement methods or other template codes, etc.
Recently, introducing Machine Learning (ML) techniques into program repair has also attracted a lot of interest and became a trend~\cite{allamanis2018survey, gupta2017deepfix, tufano2018empirical, chen2019sequencer}, which build generic models to capture statistical characteristics using previous code changes and automatically fixing the code being inserted.

However, although some promising results have been achieved, current studies of automated program repair face a list of limitations, especially on fixing vulnerabilities. 
Firstly, most APR approaches heavily rely on domain-specific knowledge or predefined change templates, which leads to limited scalability~\cite{monperrus2018automatic}.
Tufano's dataset~\cite{tufano2019empirical} contains 2 million sentence pairs of historical bug fix records.
Nevertheless, a vulnerability fix dataset such as Ponta's dataset~\cite{ponta2019manually} and the AOSP dataset ~\cite{aospdataset} only contain 624 and 1380 publicly disclosed vulnerabilities.
The confirmed CVE records number is nearly 150K \footnote{https://cve.mitre.org/}.
This means we need to train and learn from a small dataset of vulnerabilities. 
Secondly, traditional techniques leverage search space exploration, statistical analysis to rank similar repair records~\cite{goues2019automated}. 
These techniques need to define large numbers of features, which can be time-consuming and not accurate enough.
ML models can alleviate these problems but as mentioned above, only a few studies have been done to focus on vulnerability fixing because of the small sample size.

In this paper, we focus on the two issues raised above and rely entirely on machine learning to capture grammatical and structural information as common change patterns. 
In order to solve the small sample size problem, we use the fine-tuning method~\cite{tajbakhsh2016convolutional}.
Fine-tuning means that if our specialized domain dataset is similar to the general domain dataset, we can take weights of a trained neural network and use it as initialization for a new model being trained on data from the same domain.
It has been widely utilized to speed up the training and overcome the small sample size.  
Using this method, we can combine two related works together: vulnerability fixing and bug repair.
We will first pre-train the model based on the large and diverse dataset from bug repair records to capture universal features.
Then, we will fine-tune the model on our minor vulnerability fixing dataset, freeze or optimize some of the pre-trained weights to make the model more suitable for vulnerability fixing work.

We choose the general approach of Neural Machine Translation (NMT) to learn rules from historical records and apply them to future edits. 
It is widely utilized in Natural Language Processing (NLP) domain, such as translating one language (e.g., English) to another language (e.g., Swedish). 
The NMT model can generalize numerous sequence pairs between two languages, learn the probability distribution of changes, and assign higher weights to appropriate editing operations. 
Previous studies such as Tufano et al.~\cite{tufano2018empirical} and Chen et al.~\cite{chen2019sequencer} have shown an initial success of using the NMT model for predicting code changes. 

However, they only focus on simple scenarios such as short sequences and single-line cases. 
Since the NMT model is originally exploited for natural language processing, there is a distinction between natural language and programming language~\cite{casalnuovo2019studying}.
Firstly, program language falls under the category of language called context-sensitive languages.
Dependencies in one statement may come from the entire function or even the entire class.
Nevertheless, in natural language, token dependencies are always distributed in the same or neighboring sentences.
Secondly, the vocabulary of natural languages is filled with conceptual terms. 
The vocabulary of programming languages is generally only grammar words like essential comments, plus various custom-named things like variables and functions.
Thirdly, programming languages are unambiguous, while natural languages are often multiplied ambiguous and require interpretation in context to be fully understood.

In order to solve the dependency problem across the entire class, we construct the define-use (def-use)~\cite{shi2010use} chain, which represents the data-flow dependencies to capture important context around the vulnerable statement. 
It will extract all variable definitions from the vulnerable statements.
We use the state-of-the-art transformer model~\cite{vaswani2017attention} to reduce the performance degradation caused by long statements. 
This enables us to process long statements and captures a broader range of dependencies.

We called our approach SeqTrans, and it works as follows: 
Firstly, we collect historical bug and vulnerability fixing records from two previous open datasets, which contain 2 million and 5k sentence pairs of confirmed fix records.
Secondly, we start by training a transformer model with a self-attention mechanism~\cite{vaswani2017attention} for bug repairing on the big dataset.
Then, we fine-tune the model on the small dataset to match the target of our work for vulnerability fixing. 
Thirdly, if a new vulnerable object is inputted to the trained model, beam search~\cite{wiseman2016sequence} will be utilized first to obtain a list of candidate predictions. 
Then, a syntax checker will filter the candidate list and select the most suitable prediction.

In order to evaluate our approach, we calculate the accuracy at statement level and across the CVE on Ponta's dataset~\cite{ponta2019manually}. 
The experimental result shows that our approach SeqTrans reaches a promising accuracy of single line prediction by $23.3\%$ when Beam=50, outperforms the state-of-the-art model SequenceR~\cite{chen2019sequencer} by $5\%$ and substantially surpasses the performance Tufano et al. ~\cite{tufano2018empirical} and other NMT models. 
As for predicting the full CVE, our approach also achieves the accuracy of $25.3\%$ when Beam=50, which is also better than other approaches.
We also conducted a traditional evaluation experiment to verify our actual performance.
The result shows that among the 120 CVEs we select from 5 open-source projects, we correctly fix 21 of them.
We believe these promising results can confirm that SeqTrans is a competitive approach that achieves good performance on the task of vulnerability fixing.

In the meantime, we also made some ablation studies and observed internally what SeqTrans could well predict types of vulnerability fixes. 
An interesting observation we find is that our model gives results that vary for different types of CWEs. 
Our model performs quite well in specific types of CWEs like CWE-287 (Improper Authentication) and CWE-863 (Incorrect Authorization) but even cannot make any prediction for certain CWEs like CWE-918 (Server-Side Request Forgery). 
The conclusion is that training a general model to fix vulnerabilities automatically is too ambitious to cover all cases. 
However, if we can focus on specific types of CWEs, the NMT model can provide developers with promising results.
SeqTrans can cover about 25\% of the types of CWEs in the data set.

The paper makes the following contributions:
\begin{enumerate}
	\item We use the NMT model transformer to learn and generalize common patterns from historical data for vulnerability fixing.
	\item We propose to leverage data-flow dependencies to construct vulnerable sequences and maintain the vital context around them.
	\item Fine-tuning has been introduced to overcome the small sample size problem.
	\item We implement our approach SeqTrans and evaluate real publicly disclosed vulnerabilities on open-source Java projects. Our SeqTrans outperforms other program repair techniques and achieves the accuracy of 23.3\% in statement-level validation and 25.3\% in CVE-level validation.
	\item We make an internal observation about prediction results on different CWEs and find some interesting CWE fixing operations captured by our model. Our model can predict specific types of CWEs pretty well.
\end{enumerate}

\section{Motivating Examples}
\label{sec:example}

\begin{figure*}[t]
	\centering
	\subfloat[CVE-2017-1000390, jenkinsci/tikal-multijob-plugin, 2424cec7a099fe4392f052a754fadc28de9f8d86]
	{\includegraphics[width=0.9\textwidth]{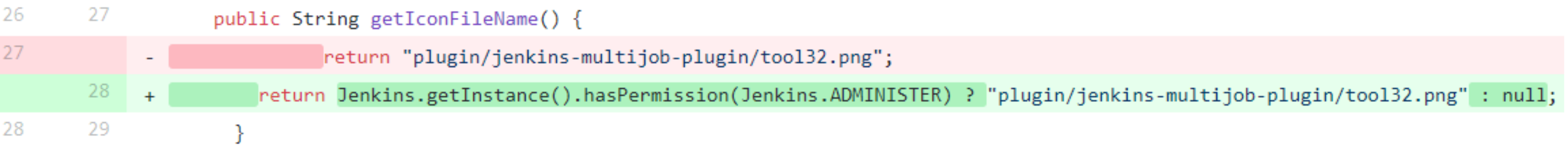}}\quad
	\subfloat[CVE-2017-1000388, jenkinsci/tikal-multijob-plugin, d442ff671965c279770b28e37dc63a6ab73c0f0e]
	{\includegraphics[width=0.9\textwidth]{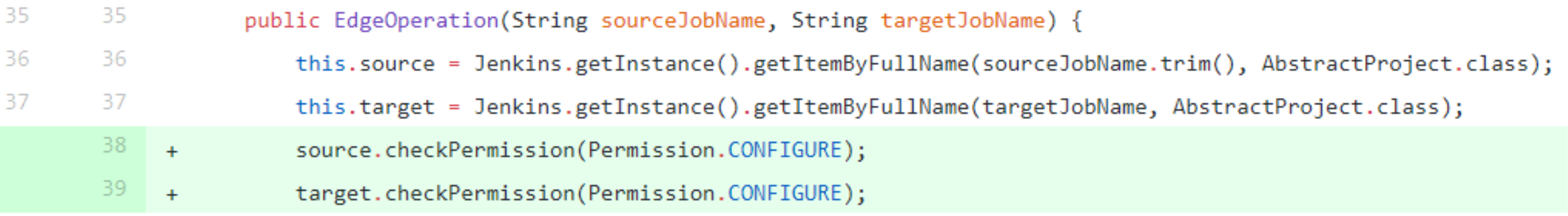}}
	\caption{Two similar vulnerability fixes belonging to CWE-732} 
	\label{fig:example}
\end{figure*}

\begin{figure*}[t]
	\centering
	\subfloat[CVE-2014-0075, apache/tomcat, f646a5acd5e32d6f5a2d9bf1d94ca66b65477675]
	{\includegraphics[width=0.45\textwidth]{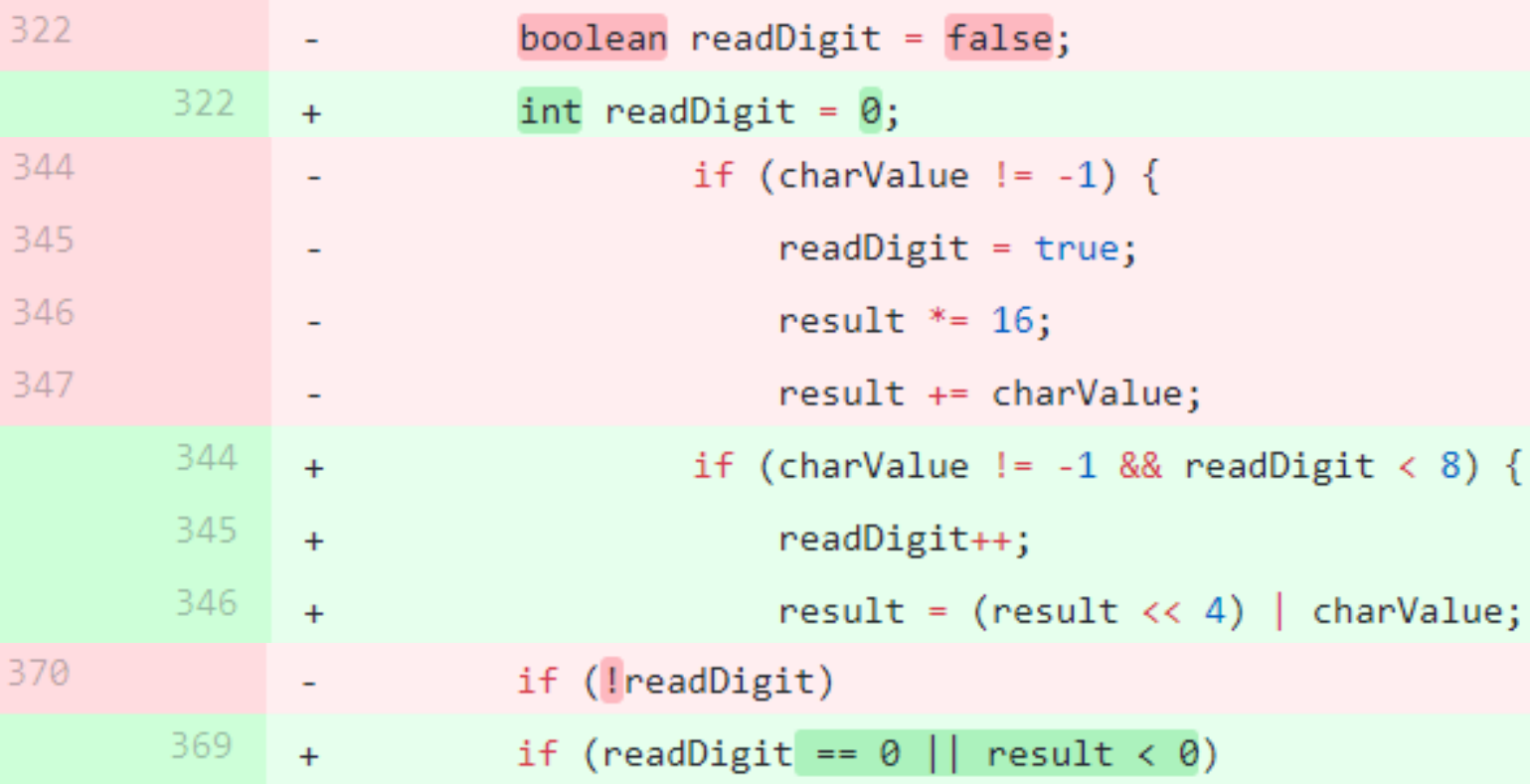}}\quad
	\subfloat[CVE-2014-0099, apache/tomcat70, 184cdc0d3f03f5737e12d21fff246d7285034597]
	{\includegraphics[width=0.45\textwidth]{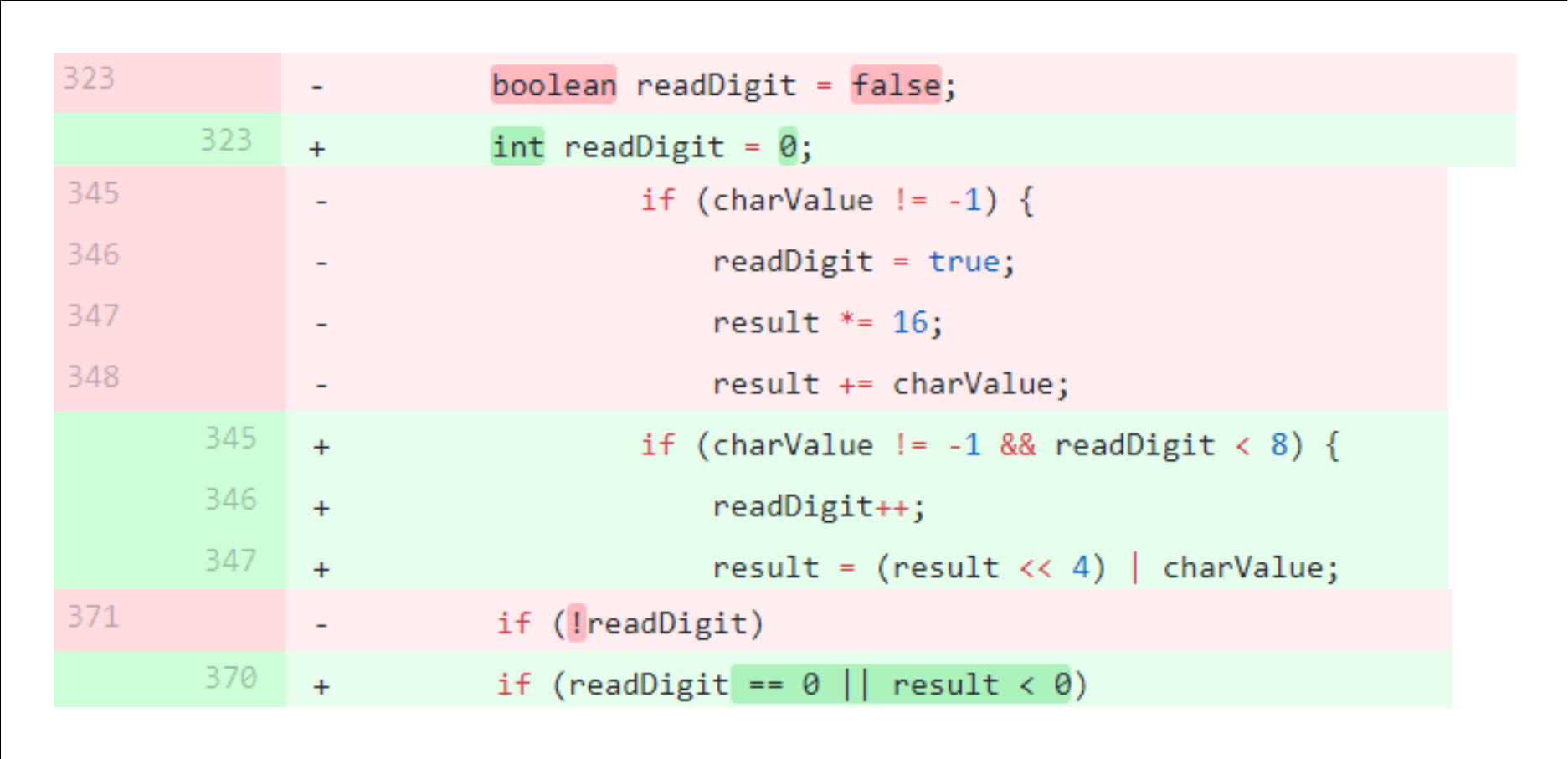}}
	\caption{Two identical vulnerability fixes belonging to CWE-189} 
	\label{fig:example2}
\end{figure*}

Figure~\ref{fig:example} shows a motivating example of our approach. In Figure~\ref{fig:example}, there are two vulnerability fixes for CVE-2017-1000390 and CVE-2017-1000388, respectively. 
These two CVEs belong to the same CWE: CWE-732, which is named "Incorrect Permission Assignment for Critical Resource". 
CWE-732 emphasizes that "the product specifies permissions for a security-critical resource in a way that allows that resource to be read or modified by unintended actors", which means that when using a critical resource such as a configuration file, the program should carefully check if the resource has insecure permissions.

In Figure~\ref{fig:example} (a), before the function \texttt{getIconFileName} returns the $ IconFileName $, it should check whether the user has the corresponding permission. 
A similar vulnerability is included in Figure~\ref{fig:example} (b). 
Before the function \texttt{EdgeOperation} accesses two resources $ JobName $, it should first confirm whether the user has the permission. 
Otherwise, it will constitute an out-of-bounds permission, which can lead to the leakage of sensitive data such as privacy.
Although these two CVEs belong to different projects, their repair processes are very similar. 
This inspired us that it might be possible to learn common patterns from historical vulnerability fixes that correspond to the same or similar CWEs.

Figure~\ref{fig:example2} is a more extreme situation, containing two identical CVE modifications CVE-2014-0075 and CVE-2014-0099. 
These two CVEs belong to the same CWE-189, which is named "Numeric Errors".
This CWE is easy to understand. 
Weaknesses in this category are related to improper calculation or conversion of numbers.
These two CVEs contain a series of modifications for overflow evasion, and they are identical.
We can directly copy the experience learned in one project to another project.

In this paper, we proposed a novel method to exploit historical vulnerability fix records to provide suggestions and automatically fix the source code.
If the function with similar structure requests accesses to a critical resource, our deep learning model can learn to check permissions before allowing access, eliminating the tedious process for developers to search for vulnerability and recapitulate repair patterns.
\section{Background}

Before describing our approach, we need to briefly introduce the transformer and other tools used in our approach.

\textbf{Tansformer}: In this work, we choose to use the transformer model~\cite{vaswani2017attention} to solve the performance degradation problem of the seq2seq model on long sequences.
It has been widely used by OpenAI and DeepMind in their language models.
The implementation of the transformer model comes from an open-source NMT framework OpenNMT~\cite{klein-etal-2017-opennmt}. 
It is designed to be research-friendly to try out new ideas in translation, summary, morphology, and many other domains. 
Some companies have proven the code to be production-ready.

Unlike Recurrent Neural Network (RNN)~\cite{mikolov2010recurrent} or Long Short Term Memory (LSTM)~\cite{gers2000learning} models, transformer relies entirely on the self-attention mechanism to draw global dependencies between input and output data. 
This model is more parallel and achieves better translation results.
The transformer consists of two main components: a set of encoders chained together and a set of decoders chained together. 
The encode-decoder structure is widely used in NMT models, the encoder maps an input sequence of symbol representations $ (x_1, ..., x_n) $ to an embedding representation $ z = (z_1, ..., z_n) $, which contains information about the parts of the inputs which are relevant to each other. 
Given $ z $, the decoder then exploits this incorporated contextual information to generate an output sequence. Generates an output sequence $ (y_1, ..., y_m) $ of symbols one element at a time. 
At each step, the model consumes the previously generated symbols as additional input when generating the next~\cite{graves2013generating}.
The transformer follows this overall architecture using stacked self-attention and point-wise, fully connected layers for both the encoder and decoder.
Each encoder and decoder make use of an attention mechanism to weigh the connections between every input and refer to that information to generate output~\cite{vaswani2017attention}.
The key design of the transformer that brings the biggest performance improvement is to set the distance between any two words to 1, which is very effective in solving the tricky long-term dependency problem in NLP~\cite{vaswani2017attention}.

\textbf{Fine-tuning}: Fine-tuning means taking weights of a trained neural network and using it as initialization or a fixed feature extractor for the task of interest~\cite{tajbakhsh2016convolutional}.
Why do we need to fine-tune?
The reasons are shown as follows~\cite{finetunestanford}:
\begin{enumerate}
	\item Overcome small sample size: it is impractical to train a large neural network, and overfitting cannot be avoided. At this time, if we still want to use the super feature extraction ability of large neural networks, we can only rely on fine-tuning the already trained models.
	\item Low training costs in the later stages: it can reduce training costs and speed up training.
	\item No need to build the wheel over and over again: the model trained by the previous work with great effort will be stronger than the model built from scratch in a large probability.	
\end{enumerate}
Using this method, we can combine two related works, such as vulnerability fixing and bug repair.
The process of fine-tuning usually consists of three parts~\cite{finetunestanford}:
\begin{enumerate}
	\item Pre-train a neural network model on the source dataset.
	\item Create a new neural network target model. It replicates all the model designs and their parameters on the source model except for the last output layer. 
	\item Train the target model on the target dataset. We will train the output layer from scratch.
\end{enumerate}

\textbf{Gumtree}: GumTree is the state-of-the-art diff searching tool~\cite{falleri2014fine}. 
It provides several interfaces to accommodate different kinds of parsers such as srcML~\cite{srcml} to parse the source code and build the AST tree.
It is worth noting that GumTree only provides a fine-grained mapping between AST nodes, so we modified the code of GumTree and combined it with another tool, Understand~\cite{understand}, to extract the precise diffs.
In the meantime, we found some bugs in Gumtree that led to incorrect mismatching and reported them to the author.
These issues are explained in more detail in Section 6.2.
The algorithm of Gumtree is inspired by the way developers manually look at changes between files. 
It will traverse the AST tree pairs and compute the mappings in two successive phases:
\begin{enumerate}
	\item A greedy top-down algorithm to find isomorphic sub-trees of decreasing height. Mappings are established between the nodes of these isomorphic subtrees. They are called anchors mappings.
	\item A bottom-up algorithm where two nodes match (called a container mapping) if their descendants (children of the nodes, and their children, and so on) include a large number of common anchors. When two nodes match, an optimal algorithm will be applied to search for additional mappings (called recovery mappings) among their descendants. 
\end{enumerate}

\section{Methods}
\label{sec:ours}

\begin{figure*}
	\centering
	\includegraphics[width=500pt]{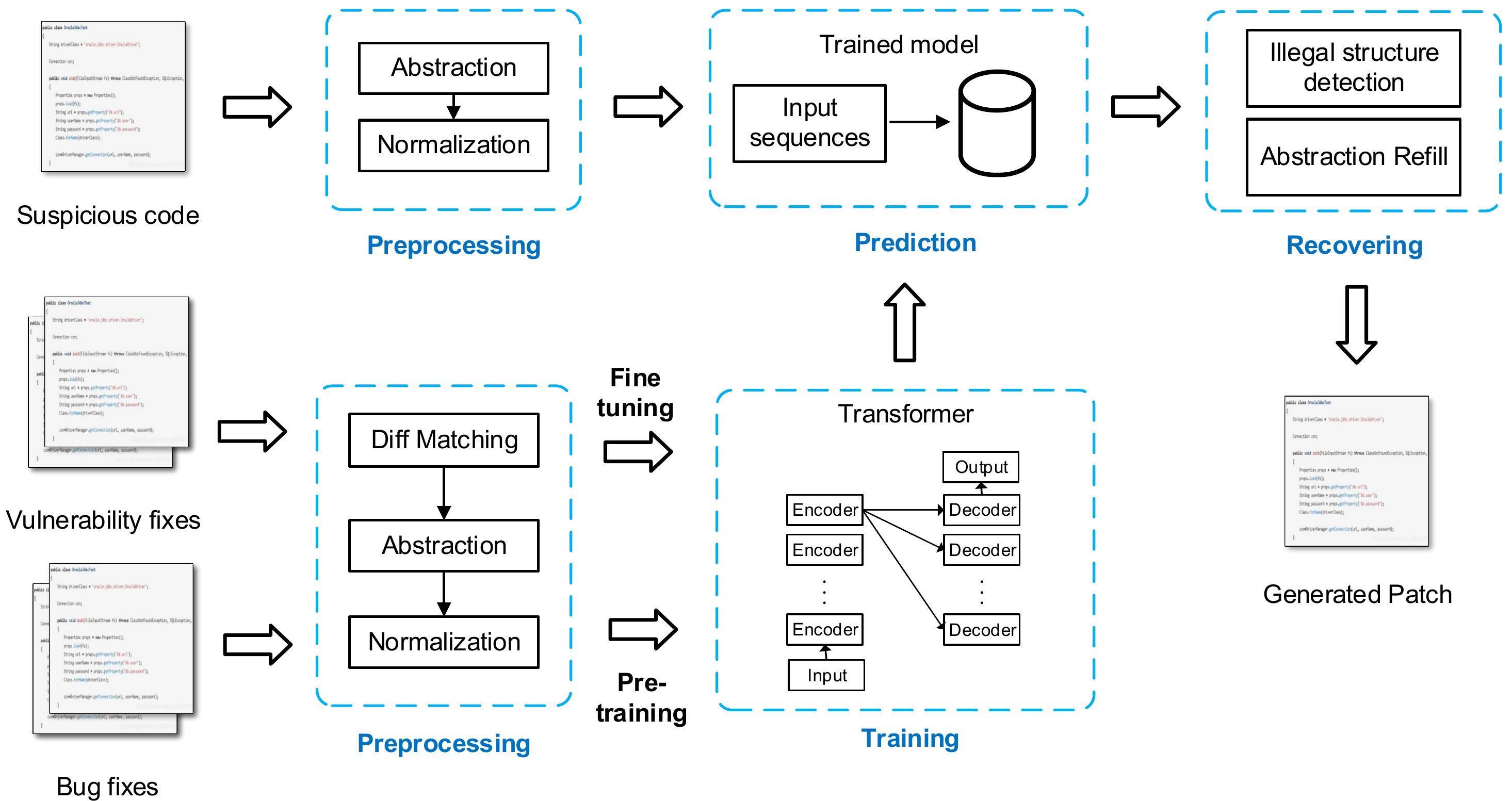}
	\caption{Overview of our SeqTrans for automatically vulnerability fixing}
	\label{fig:overview}
\end{figure*}

We use the NMT method to automatically guide vulnerability fixing, which aims to learn common change patterns from historical records and apply them to the new input files. 
In order to overcome the small sample size problem, we introduce the fine-tuning technique.
Data-flow dependencies have also been introduced to maintain and capture more critical information around the diff context.
SeqTrans can work together with other vulnerability detection tools such as Eclipse Steady~\cite{ponta2018beyond}. They can provide vulnerability location information at the method level.

\subsection{Overview}
The overview of our approach is given in Figure ~\ref{fig:overview}, which contains three stages: preprocessing, pre-training and fine-tuning, prediction and patching.

\textbf{Preprocessing:} In this step, we will extract diff contexts from two datasets: bug repair and vulnerability fixing datasets.
Then, we perform normalization and abstraction based on data-flow dependencies to extract the def-use chains. 
We believe def-use chains are suitable for deep learning models to capture syntax and structure information around the vulnerabilities with fewer noises. 
These def-use chains can be fed into the transformer model.

\textbf{Pre-training and fine-tuning:} The training process starts on the bug repair dataset because bug repairs are easier to collect a large enough training set than vulnerability fixes.
The tasks of vulnerability and bug fixing have something in common, in other words, something to learn from each other.
We can learn and capture parts of general features and hyperparameters from the general task domain dataset, the bug repair dataset.
After the pre-training, we will fine-tune the transformer model on the vulnerability fixing dataset. 
This dataset is much smaller than the first dataset because it is hard to confirm and collect a big enough size for training.
Based on the first model, we will refine some of the weights to make the model more suitable for the task of vulnerability fixing.
Fine-tuning has been proven to achieve better results on small datasets and speeds up the training process~\cite{mahajan2018exploring, liu2019roberta}.

\textbf{Prediction and patching:}
If one vulnerable file is inputted, we need to locate the suspicious codes and predict based on the trained model.
In this paper, we do not pay much attention to the vulnerability location part. 
They can be accomplished by previous vulnerability location tools or with the help of a human security specialist.
SeqTrans can provide multiple candidates for users to select the most suitable prediction.
Syntax checker Findbugs~\cite{findbugs} is exploited to check for errors and filter out predictions that contain syntax errors in advance.
After that, we refill abstraction and generate patches. We will discuss the details of each part in the following part of this section.

\subsection{Code Change Mining}
The two datasets we utilized are Tufano's~\cite{tufano2019empirical} and Ponta's datasets~\cite{ponta2019manually}.
Tufano's dataset provides raw source code pairs extracted from the bug-fixing commits, which is easy to be used. 
However, Ponta's dataset only provides the CSV table containing the vulnerability fixing records.
We need a crawler to crawl the project we want.
The table contains vulnerability fixing records are shown as follows:
\begin{displaymath}
	(vulnerability\_id; repository\_url; commit\_id)
\end{displaymath}
where $ vulnerability\_id $ is the identifier of a vulnerability fixed in the $ commit\_id $ in the open-source code repository at the $ repository\_url $.
Each line in the dataset represents a commit that contributes to fixing a vulnerability.
Then, we utilize a crawler to collect program repositories mentioned in the dataset. 
Pull Request (PR) data will be extracted based on $ commit\_id $.
After that, we need to find out Java file changes involved in each PR. 
Because our approach SeqTrans only supports Java files now.
With the help of a git version control system JGit~\cite{jgit}, we can retrieve the version of Java files before and after code changes implemented in the PR. 
We call these Java file pairs $ Change Pair (CP) $, each $ CP $ contains a list of code diffs.
In some cases, repair operations are performed only on XML or other resource files, or the entire file is refactored directly.
In these cases, examples are filtered out.
Lastly, we extracted 5K and 650K $CPs$ from Ponta's and Tufano's datasets.

\subsection{Code Diff Extraction}
\label{sec:diff_extra}
After obtaining $ CPs $ from PR, we need to locate the diff context. 
Although we can exploit the \textit{"git diff"} command provided by git to search line-level code diffs, it just does not fulfill our needs. 
Slight code structure changes such as a new line and adding space are not required.
For this reason, we choose to search for code diffs by using Abstract Syntax Trees (ASTs).
The state-of-the-art diff searching tool named GumTree~\cite{falleri2014fine} is utilized to search for fine-grained AST node mappings.

After that, each CP is represented as a list of code diffs:
\begin{displaymath}
	CP = {(st_{src}, st_{dst})_1,...,(st_{src}, st_{dst})_n}
\end{displaymath}
where $ (st_{src}, st_{dst}) $ represents statements from the source file and the destination file. 

\begin{figure}[!t]
	\centering
	\includegraphics[width=2in]{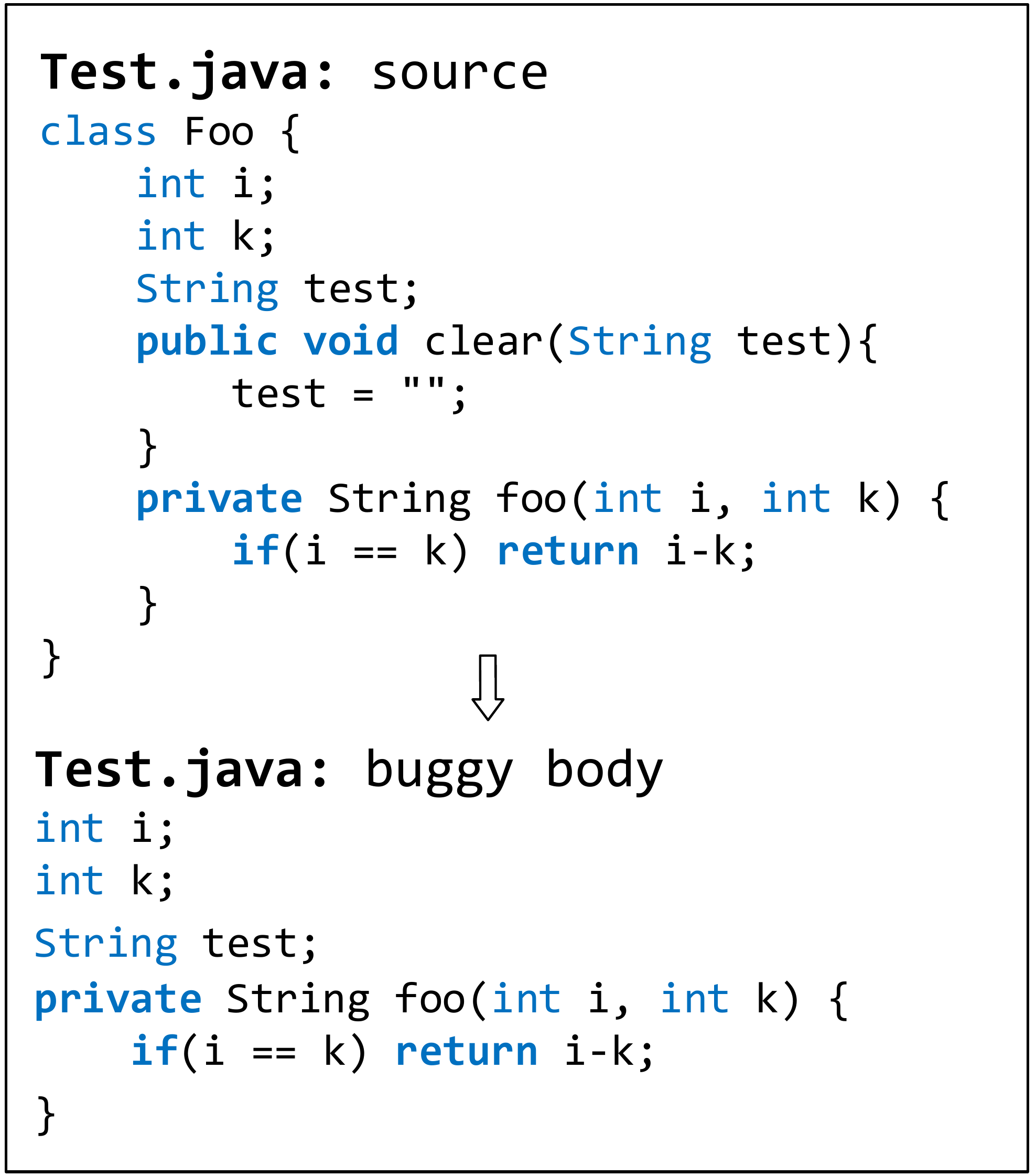}
	\caption{One example of the buggy body}
	\label{fig:buggy_code}
\end{figure}

Then, we will extract data-flow dependencies around code diffs to construct our def-use chains. 
A def-use chain means assigning some value to a variable, containing all variable definitions from the vulnerable statement. 
The reasons why we use data-flow dependencies are shown as follows: 
1) Context around the vulnerable statements is valuable for understanding risky behavior and capturing structure relationships.  
However, it is too heavy to maintain the full context with lots of unrelated code at the class level. 
2) Data-flow dependencies provide enough context for transformation. 
If one statement needs to be modified, it is highly likely to co-change its definitions simultaneously. 
3) Control flow dependencies often contain branches, making them too long to be tokenized.
One example has been given in Figure \ref{fig:buggy_code}. 
Assume that the method "foo" contains one vulnerability, we will maintain the method and the vulnerable statement.
All global variables will be preserved.
All statements that have data dependencies on the vulnerable statement will be retained, too.
Statements located after the vulnerable statement within the same method will be removed.

The definition and use (def-use) dependencies can be extracted from the ASTs. 
The process can be divided into three parts:
\begin{enumerate}
	\item Traverse the whole AST and label each variable name, constant name, and string name. These names are distributed over the leaf nodes of the AST. This step will be done in the first phase of the modified Gumtree algorithm.
	\item Traverse up from the leaf node to search for the defined parent nodes, record the locations.
	\item Locate the relevant definition statements of the error-prone statements by location records.
\end{enumerate}
We implement this by modifying the code of Gumtree. Another static analysis tool named Understand is also used to transfer the location records to codes.
SeqTrans will change each CP as the following shows:
\begin{displaymath}
	CP = {((def_1,..., def_n, st_{src}), (def_2,..., def_m, st_{dst}))_1,...,}
\end{displaymath}
\begin{displaymath}
	{((def_1,..., def_n, st_{src}), (def_2,..., def_m, st_{dst}))_n}
\end{displaymath}
In this paper, we ignore code changes that involve the addition or deletion of entire methods/files.

\subsection{Normalization \& Tokenization}
In the training process of the NMT model, there exist a couple of drawbacks. 
Because NMT models output a probability distribution over words, they can become very slow with many possible words. 
We need to impose an artificial limit on how many of the most common words we want our model to handle. 
This is also called vocabulary size.
In order to reduce the vocabulary size, we need to preserve the semantic information of the source code while abstracting the context.

\begin{figure}[!t]
	\centering
	\includegraphics[width=2in]{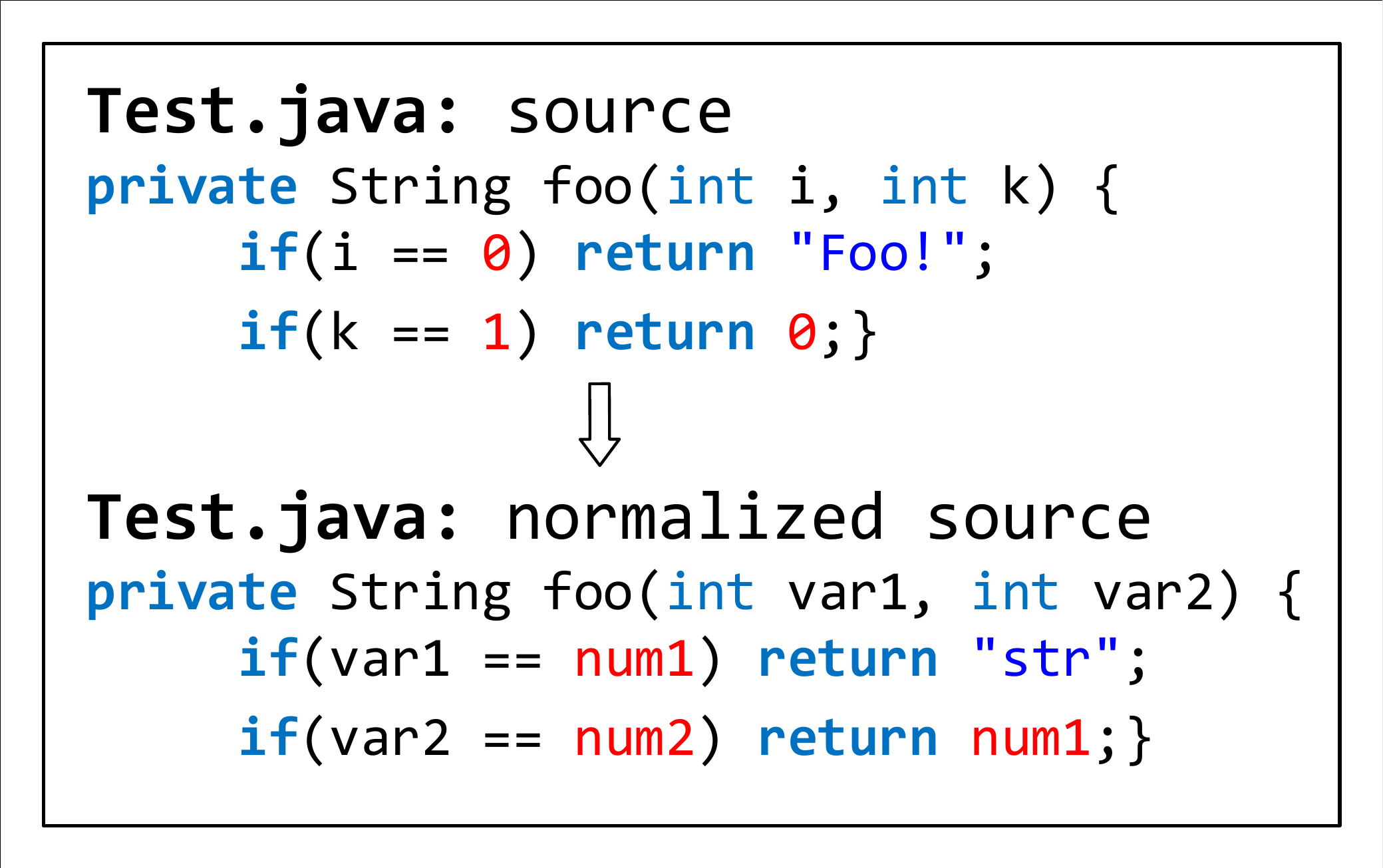}
	\caption{Normalize the source code}
	\label{fig:normal}
\end{figure}
The normalization process is shown in Figure~\ref{fig:normal}. We replace variable names to $ ``var1",\ldots,``varn" $, each literal and string are also replaced to $ "num1",\ldots,"numn" $ and $ "liter" $. 
The reasons why we do this involve: 1) reducing the vocabulary size and the frequency of specific tokens; 2) reducing the redundancy of the data and improving the consistency of the data.
We will maintain a dictionary to store the mappings between the original label and the substitute to be refilled after prediction.
We can control the vocabulary size and make the NMT model concentrate on learning common patterns from different code changes through the above optimization. 


Subsequently, we split each abstract CP into a series of tokens. 
It is worth mentioning that the seq2seq model utilized in previous studies faces severe performance degradation when processing long sequences.
For example, Tufano et al.\cite{tufano2018empirical} limited the token number to 50-100.
By utilizing the transformer model we can better handle long sequences. 
In our approach, we will limit the CP to 1500 tokens.
The vocabulary size is set to 8k based on Gowda's work~\cite{gowda2020finding}.
We will discuss the details in the following sections.

\subsection{Neural Machine Translation Network}
In this phase, we train SeqTrans to transform the vulnerable codes and generate multiple prediction candidates.
The training process can be divided into two phases: pre-training and fine-tuning.

\subsubsection{Pre-training}
In the pre-training process, we will utilize a generalized domain corpus from Tufano's dataset for bug repairing to perform the first training.
Vulnerability fixing can be considered as a subset of bug repairing. 
We believe that by pre-training on generic data, we can learn many generic fixing experiences and features that can be applied to the task of vulnerability fixing.
A list of $ CPs_{general} $ will be extracted by using the approach discussed in section~\ref{sec:diff_extra}.
These $ CPs_{general} $ that contain vulnerable version and fixed version diff context will be given to the network.
We will discuss the network in detail in the following subsection.
The pre-training model will be trained for 300K steps till convergence because we found that the validation accuracy smoothed at this training step and no longer fluctuated.
In the next fine-tuning process, we will select the model with the highest accuracy on the validation dataset as the final model.
The model comes from a breakpoint backup every 5K steps.

\subsubsection{Fine-tuning}
The purpose of fine-tuning is to improve the model's generalization ability when the target dataset is much smaller than the source dataset.
Using this method, we can combine two related works: vulnerability fixing and bug repair.
However, one issue is that although fine-tuning is widely used in the Neural Language (NL) field and many pre-training models are provided, there are very few such pre-trained models in the Programming language (PL) field.
That is why we need to train the generic domain model by ourselves.
The model trained in the previous training process will be fine-tuned using a new vulnerability fixing corpus so that the knowledge learned in the bug repair training can be transferred to the vulnerability fixing task.
We set the step size to 1/10 of the pre-training step size.
The model selection process is the same as the previous step.

Due to overfitting concerns~\cite{thompson2018freezing}, we will keep earlier layers fixed and only fine-tune the last layer of the model.
The training process will update the vocabulary corpus and continue till convergence.
A smaller learning rate was selected than the pre-training process, which was set to 0.01.
It is worth noting that some studies such as Gururangan's work~\cite{gururangan2020don} and documents of OpenNMT\cite{opennmt} mentioned that some sequences were translated poorly (like unidiomatic structure or UNKs) by the retrained model while they are translated better by the base model, which is called "Catastrophic Forgetting".
In order to alleviate the catastrophic forgetting, the retraining should be a combination of in-domain and generic data.
In this work, we will try to mix part of general domain data into specific domain data to generate such a combination.
We have roughly selected some data to be blended into the special domain data on the basis that the blended data should not expand the size of the corpus as much as possible.
Eventually, we will double the size of the training set, and the test set will remain unchanged.

\subsubsection{Encoder} The encoder is composed of a stack of 6 identical layers.
Each layer consists of two sub-layers: a multi-head self-attention mechanism and a feed-forward neural network. 
Residual connection~\cite{he2016deep} and normalization~\cite{ba2016layer} have been employed to each sub-layer so that we can represent the output of the sub-layer as:
\begin{displaymath}
	sub\_layer\_output = Layer\_normization(x+(SubLayer(x)))
\end{displaymath}
where $ Sublayer(x) $ is the function implemented by the sub-layer itself.
The self-attention mechanism takes in a set of input encodings from the previous encoder and weighs their relevance to each other to generate a set of output encodings. 
The feed-forward neural network then further processes each output encoding individually. 
These output encodings are finally passed to the next encoder as its input. 
The padding mask has been utilized to ensure that the encoder does not pay any attention to padding tokens.
All sub-layers as well as the embedding layers produce outputs of dimension $ d_{model} = 512 $

\subsubsection{Decoder} The decoder also contains a stack of 6 identical layers.
However, each layer consists of three sub-layers: an attention sub-layer has been added to perform multi-head attention to draw relevant information from the encodings generated by the encoders.
The masking mechanism that contains padding mask and sequence mask has been used to prevent positions from attending to subsequent positions and ensure that the predictions for position $ i $ can depend only on the known outputs at positions less than $ i $~\cite{vaswani2017attention}.
The other parts are the same as the encoder.

\subsubsection{Attention Mechanism} 
The purpose of an attention mechanism is to use a set of encodings to incorporate context into a sequence.
For each token the attention mechanism requires a query vector $ Q $ of dimension $ d_k $, a key vector $ K $ of dimension $ d_k $ and a value vector $ V $ of dimension $ d_v $. 
These vectors are created by multiplying the embedding by three matrices trained during the training process.
The essence of the attention mechanism is actually an addressing process, which is the embodiment of the attention mechanism to alleviate the complexity of the neural network model: instead of feeding all $ N $ inputs to the neural network for computation, only some task-relevant information from $ X $ needs to be selected and fed to the neural network.
Self-attention refers to the situation where the queries, keys, and values are all created using sequence encodings.
Then the output $ Z $ of this attention mechanism is:
\begin{displaymath}
	Z = Attention(Q, K, V) = softmax(\frac{QK^T}{\sqrt{n}})V
\end{displaymath}
The multi-head attention utilized in the transformer implements several attention mechanisms in parallel and then combines the resulting encoding in a process.

\begin{figure*}
	\centering
	\includegraphics[width=0.8\textwidth]{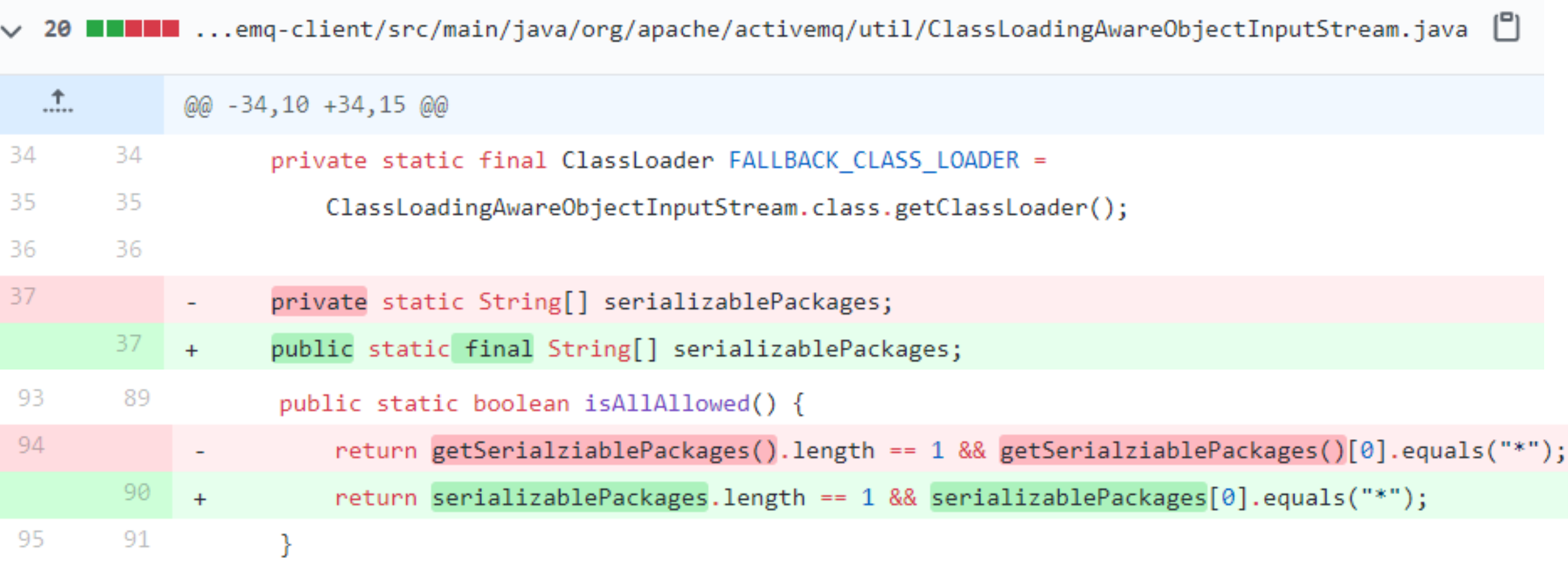}
	\caption{CVE-2015-5254, activemq, 73a0caf758f9e4916783a205c7e422b4db27905c}
	\label{fig:activemq}
\end{figure*}

\begin{figure*}
	\centering
	\includegraphics[width=0.8\textwidth]{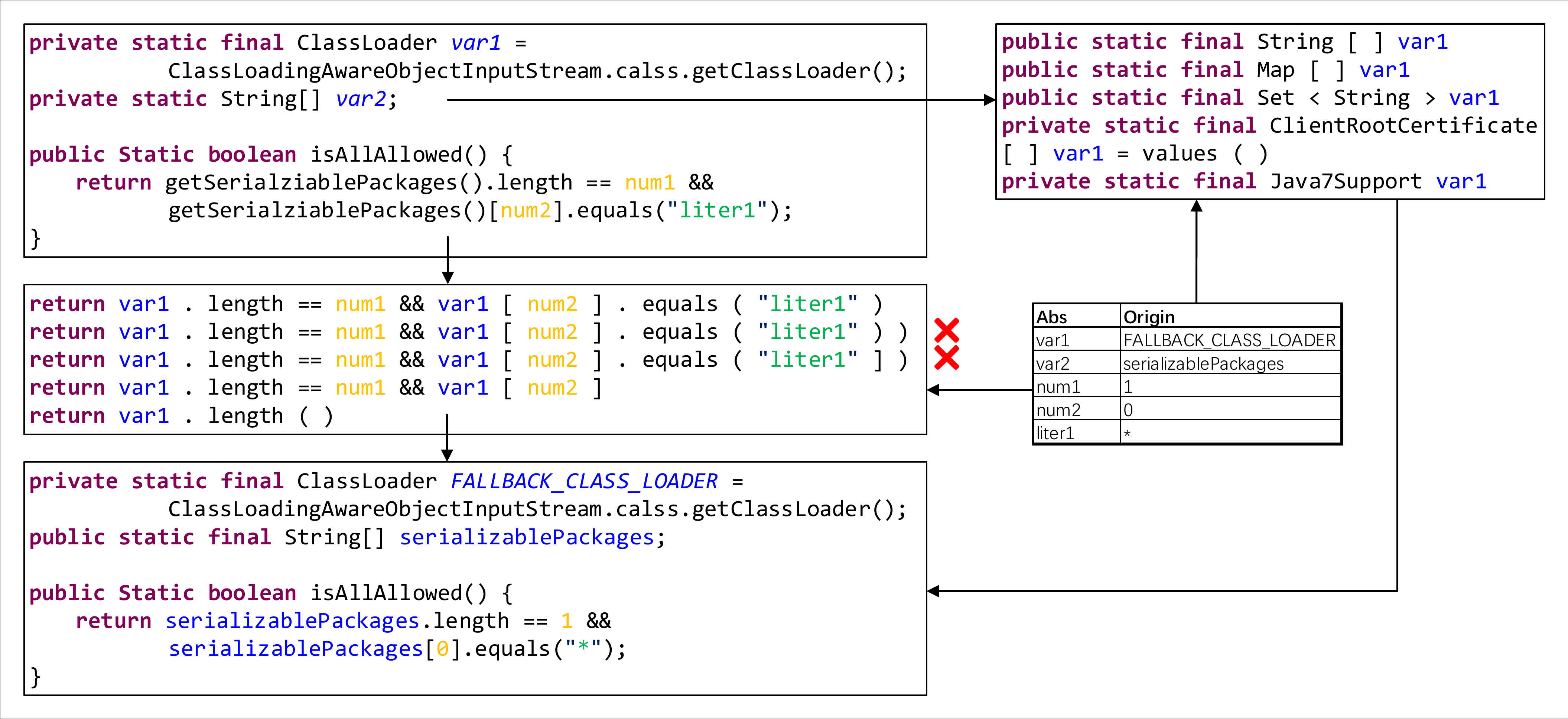}
	\caption{CVE-2015-5254, activemq, 73a0caf758f9e4916783a205c7e422b4db27905c}
	\label{fig:activemq_prediction}
\end{figure*}

\subsection{Prediction and Patch Generation}
The original output (or a list of outputs) is far from the version that can be successfully compiled. Because it contains abstraction and normalization, it even may contain grammatical errors after prediction.
Our patch generation consists of two steps to solve these problems: abstraction refill and syntax check.
In this work, we assume perfect vulnerability localization because different studies may choose different fault localization algorithms, implementations, and granularities such as method-level or statement-level. 
Liu et al. has pointed out that it is hard to compare different repair techniques due to the reason of different assumptions about the fault localization~\cite{liu2019you}.
We have made a discussion about fault localization in Section~\ref{sec:discussion}.
Vulnerable codes can come from a classifier, a vulnerability detection tool, or suspicious codes.
We will utilize an example from the open-source project called \textit{activemq} to illustrate the process of patch inference and generation.

Figure~\ref{fig:activemq} shows a CVE repair record in \textit{activemq}, which contains three statement fixes.
Firstly, as mentioned in Figure~\ref{fig:overview}, the input codes need to be abstracted and normalized.
We decompose them into sequences following a similar process as depicted in Figure~\ref{fig:activemq_prediction}.
In Figure~\ref{fig:activemq_prediction}, every abstracted variable has been marked in blue color, with every constant in yellow color and every literal in green color.
Each sequence will maintain a dictionary for future recovery.
The location of the sequence will also be recorded for subsequent backfill.
Then, these sequences are fed into the transformer model, beam search~\cite{tufano2018empirical} are used to generate multiple predictions for the same vulnerable statement.
The output of the network is also abstracted sequences like Figure~\ref{fig:activemq_prediction}.
It is a sequence that contains the predicted statement and the context around it. 
Thirdly, we backfill all the abstractions when a prediction is selected and apply syntax checks.
The next subsections will supplement some concrete techniques and tools applied in this process.

\subsubsection{Beam Search} 
In many cases, developers have specific domain-specific knowledge.
We can generate a list of prediction results to let developers pick the most suitable one. 
Beam search a heuristic graph search algorithm~\cite{raychev2014code, freitag2017beam}.
Instead of greedily choosing the most likely next step as the sequence is constructed, the beam search expands all possible next steps and keeps the $ k $ most likely, where $ k $ is a user-specified parameter and controls the number of beams or parallel searches through the sequence of probabilities. 
Beam search maintains the n best sequences until the upper limit of the set beam size.

As has been depicted in Figure~\ref{fig:activemq_prediction}, each of the vulnerable statements will generate five prediction candidates.
Usually, the highest-ranked predictions will be chosen and utilized.
In some cases, there are syntax errors in the prediction results.
We will use syntax checking tools to detect these errors.
This will be discussed in the following subsections
These $ k $ candidates will be provided as suggestions to developers to select the best result.

\subsubsection{Abstraction Refill}
As has been shown in Figure~\ref{fig:activemq_prediction}, SeqTrans will maintain a dictionary to store the necessary information for restoration before abstraction.
After prediction, the output will be concretized, and all the abstractions in the dictionary will be refilled.
The code will be automatically indented in this process.
It should be noted that all comments will be deleted and will not be refilled again.
The dictionary we maintain will store relevant variable, constant and literal for the whole $ CP $. 
We believe that the search space explosion is not an important issue at this scale.
One shortcoming of SeqTrans is that the mappings included in the dictionary come from the source files.
If the abstraction is the content that needs to be repaired, it is hard for SeqTrans to understand and infer them.
All we can do is reduce the corresponding abstraction according to the dictionary.
For example, if one \textit{println} function changes what it wants to print.
The model has difficulty predicting what it wants to print.
If a predicted abstraction cannot find a mapping in the dictionary, we will copy the original abstraction content to the current location.

\subsubsection{Syntax Check}
We combine beam search with a grammar check tool to analyze the syntax and grammatical errors contained in the predictions.
The static analysis tool \textit{FindBugs}~\cite{findbugs} is exploited to identify different potential bugs in Java programs.
The version we utilized is 3.0.1. 
The motivation for introducing static analysis is to filter out as many invalid generation patches as possible before executing test cases. 
Because the time cost of running all the test cases is very high.
Potential errors can be divided into four levels: scariest, scary, troubling, and of concern based on their possible impact or severity.

In SeqTrans, one generated patch needs to pass the compiler first and then the FindBugs detection.
If the candidate prediction cannot pass the checking process, it will be filtered.
It should be noted that Findbugs may trigger a warning even on the pre-commit version, so we only check the warning messages that are added after the prediction.
For example, in Figure~\ref{fig:activemq_prediction}, the second and the third candidates contain a syntax error, which cannot pass the check of FindBugs.
We will remove these two candidates.
In other words, we use FindBugs to check the candidates to ensure that the five candidates we recommend introduce as few new bugs as possible.
We also make an evaluation for this checker in the experimental Section.

Finally, we can generate the newly patched file and provide it to developers.
We provide flexible choices for developers to enable this feature or judge by their domain-specific knowledge.
Developers also have the flexibility to choose the predictions they need based on their own domain experience and based on our five recommended candidates.
In addition, we believe that with the continuous improvement of model training, these grammatical errors will become less and less. 
In the end, we will no longer rely on third-party grammatical error check tools.

\section{Empirical Study \& Evaluation}
\label{sec:evaluation}
In this section, we conduct our experiment on a public dataset~\cite{ponta2019manually} of vulnerability fixes and evaluate our method: SeqTrans by investigating three research questions. 

\subsection{Research Questions}

We explore the following research questions:

\begin{itemize}
	\item \textbf{RQ1:} How much effectiveness can SeqTrans provide for vulnerable code prediction?
	
	RQ1 aims to prove that the NMT-based technique is a feasible approach to learn automated code transformations, and SeqTrans outperforms other state-of-the-art techniques.
	\item \textbf{RQ2:} What are the characteristics of the ML model used that can impact the performance of SeqTrans.
	
	RQ2 will evaluate the impacts of the main components of SeqTrans on performance, such as the data structure and the transformer model.
	\item \textbf{RQ3:} How does SeqTrans perform in predicting specific types of CWEs?
	
	RQ3 will explore in-depth the prediction results and the source codes of the data set to observe whether our method performs inconsistently when predicting different kinds of CWEs.
\end{itemize}

\subsection{Experimental Design}
In this section, we discuss our experimental design for RQ1, RQ2, and RQ3.
All experiments were accomplished on a server with an Intel Xeon E5 processor, four Nvidia 3090 GPU, and 1TB RAM.

\noindent\textbf{Dataset:} Our evaluation is based on two public datasets: Tufano's~\cite{tufano2019empirical}~\footnote[1]{https://sites.google.com/view/learning-fixes/data} and Ponta's datasets~\cite{ponta2019manually}~\footnote[2]{https://github.com/SAP/vulnerability-assessment-kb}.
Tufano's dataset contains ~780,000 bug fix commits and nearly 2 million sentence pairs of historical bug fix records.
For each bug-fixing commit, they extracted the source code before and after the bug-fix using the GitHub Compare API~\cite{gitapi}.
Each bug-fixing record contains the buggy (pre-commit) and the fixed (post-commit) code. 
They discarded commits related to non-Java files and new files created in the bug-fixing commit since there would be no buggy version to learn. 
Moreover, they discarded commits impacting more than five Java files since they aim to learn focused bug fixes that are not spread across the system.

\begin{figure}[H]
	\centering
	\includegraphics[width=2.7in]{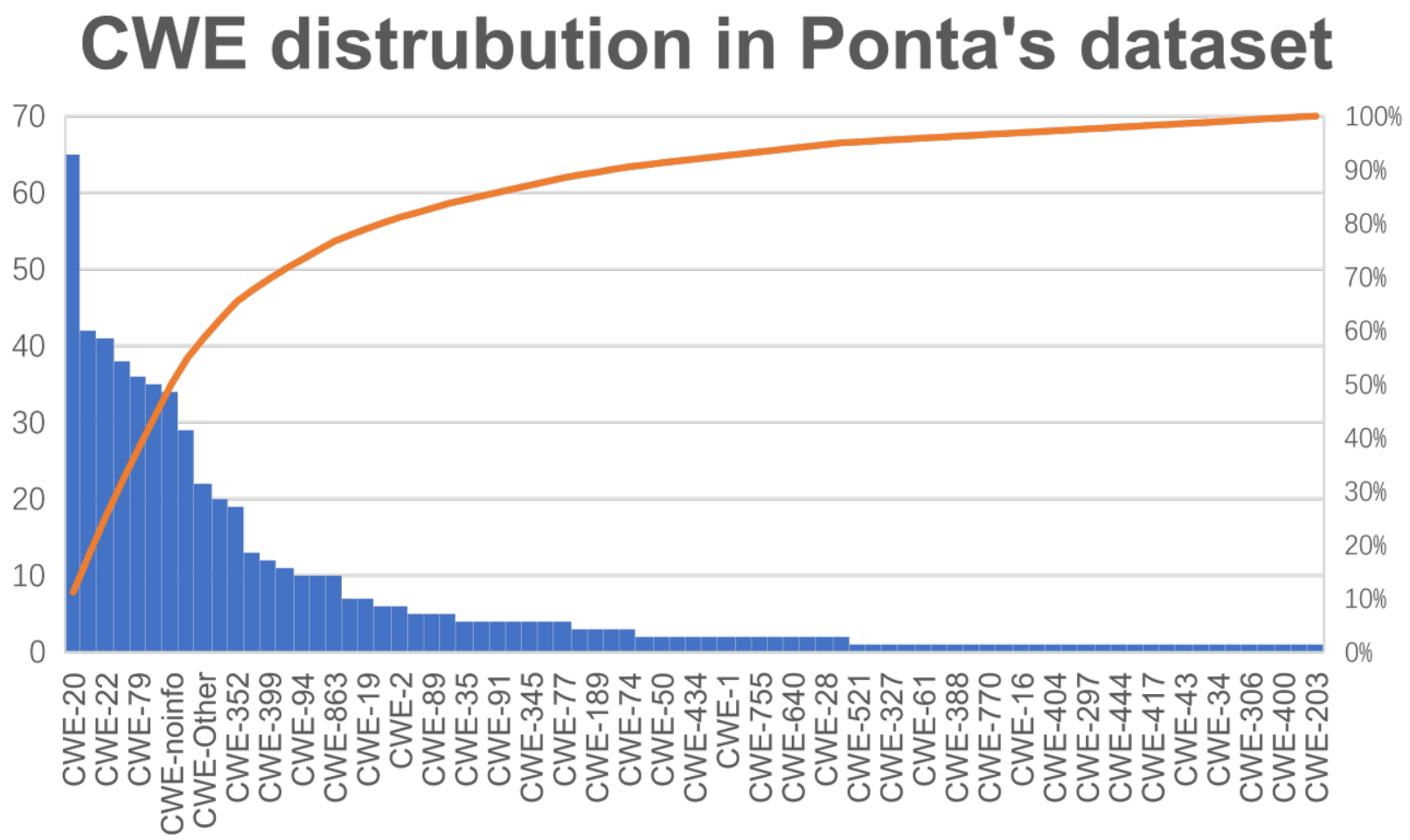}
	\caption{CWE distribution of Ponta's dataset}
	\label{fig:CWE_distribution}
\end{figure}

Ponta's dataset was obtained from the National Vulnerability Database (NVD) and from project-specific Web resources that they continuously monitor.
From that data, they extracted a dataset that maps 624 publicly disclosed vulnerabilities affecting 205 distinct open-source Java projects, used in SAP products or internal tools, onto the 1282 commits that fix them. 
The distribution of these CVEs ranges from 2008 through 2019.
Out of 624 vulnerabilities, 29 do not have a CVE identifier, and 46, which do have a CVE identifier assigned by a numbering authority, are not available in the NVD yet.
These vulnerabilities have been removed from the dataset, the final number of non-repetitive CVEs is 549 with 1068 related commits.
In total, the processed Ponta's dataset contains 1068 different vulnerabilities fixing commits with 5K diff contexts across 205 projects classified as 77 CWEs from 2008 to 2019.
Figure~\ref{fig:CWE_distribution} shows the CWE distribution in descending order of frequency, with the cumulative yellow  line on the secondary axis, identifying the percentage of the total number. 
In the appendix, we have listed the IDs and type explanations of all CWEs in Ponta's dataset.

The datasets are released under an open-source license, together with supporting scripts that allow researchers to automatically retrieve the actual content of the commits from the corresponding repositories and augment the attributes available for each instance. 
Also, these scripts complement the dataset with additional instances that are not security fixes (which is useful, for example, in machine learning applications).

\noindent \textbf{Validation:} We will use three methods to validate the performance of the experiments.

The first validation set $ T_{cross} $ is 10-fold cross-validation. 
Cross-validation is a technique to evaluate predictive models by partitioning the original sample into a training set to train the model and a test set to evaluate it.
In 10-fold cross-validation, the original sample is randomly partitioned into ten equal size subsamples. 
Of the ten subsamples, a single subsample is retained as the validation data for testing the model, and the remaining nine subsamples are used as training data. 
The process is then repeated ten times (the folds), with each of the ten subsamples used exactly once as the validation data.
It should be noted that cross-validation is only applied to fine-tuning process.
All nine subsamples will share the same pre-training set.
If the predicted statement equals the statement in the test set, there is a correct prediction.
The ten results from the folds will be averaged to produce a single estimation. 
The advantage of this method is that each sample of data is used as training data and test data. 
The over-learning and under-learning states are avoided, the results obtained are more convincing.

\begin{figure}[H]
	\centering
	\includegraphics[width=2.8in]{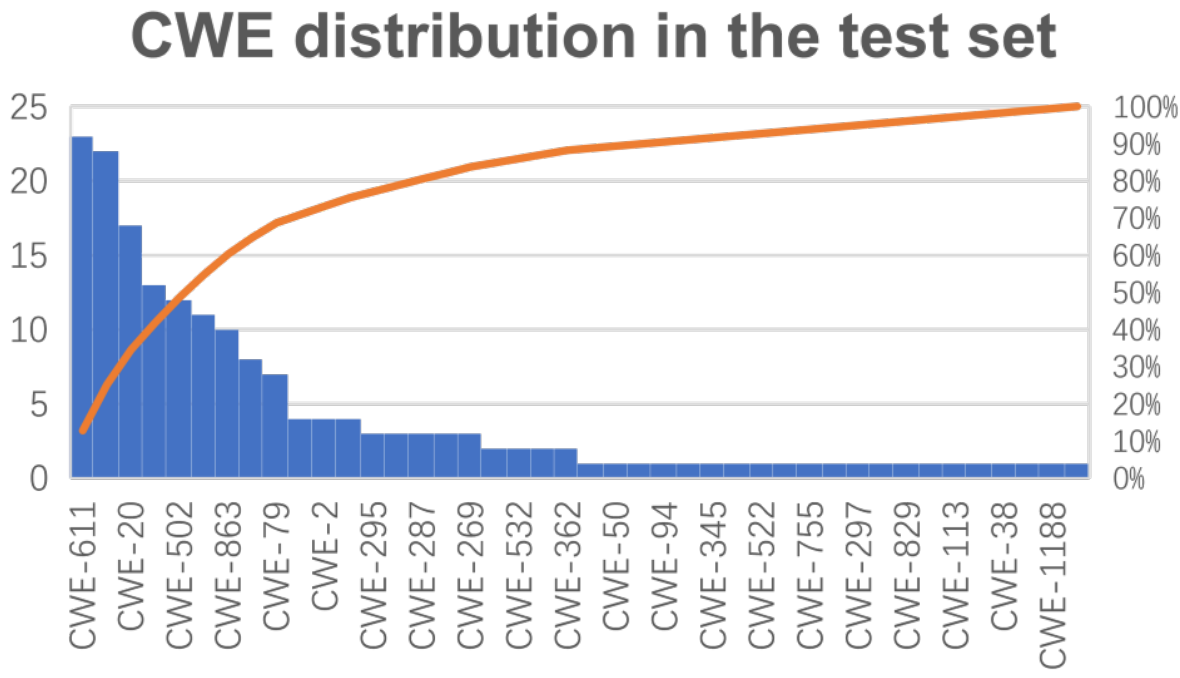}
	\caption{CWE distribution of the test set}
	\label{fig:CWE_test_distribution}
	\vspace{-0.3cm}
\end{figure}

The second validation set $ T_{cwe} $ is based on the chronological relationship of the CVE repair records to simulate the actual development process of using historical vulnerability fix records to fix subsequent suspicious code. 
We also sorted the CVE samples in Ponta's dataset by time series and used the CVE fix records from 2008 to 2017 as the training set (708 $ CPs $), the CVE fix records from 2018 and 2019 were utilized as the validation (136 $ CPs $) and test sets (150 $ CPs $).
If one $ CP $ has been fully and correctly predicted, we regard it as one successful fix.
The distribution of the 42 CWEs in the test set is shown in Figure~\ref{fig:CWE_test_distribution}.
The previous two validations do not contain compilation and syntax checker in the abstraction refill part.
We match the refilled statements strictly with the statements in historical repair records to verify if it is a correct patch.
We will verify the performance of this part in the next validation experiment.

\begin{figure}[H]
	\centering  
	\subfloat[Tufano's dataset]{
		\label{fig:dataset_single_commit}
		\includegraphics[width=0.23\textwidth]{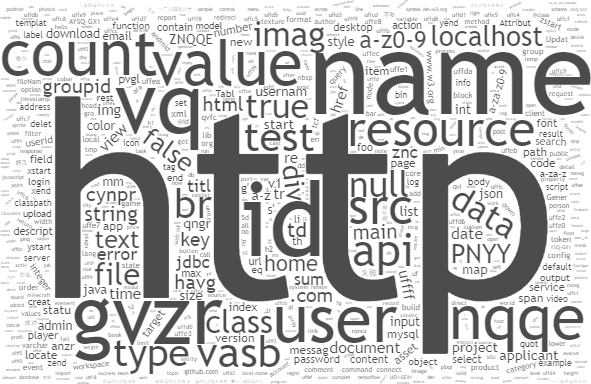}}
	\subfloat[Ponta's dataset]{
		\label{fig:dataset_single_noduplciate}
		\includegraphics[width=0.23\textwidth]{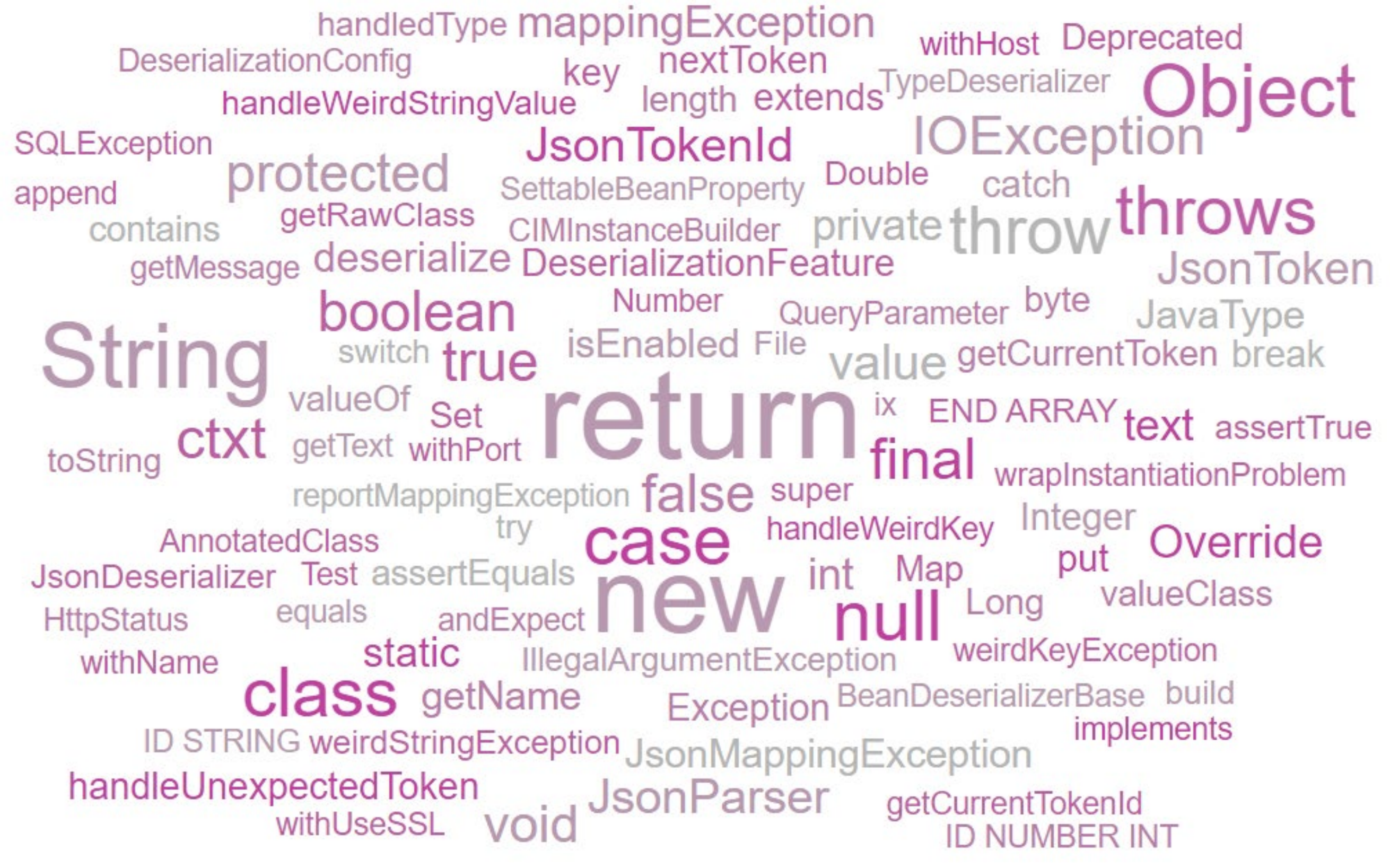}}
	\caption{Label distribution for each dataset}
	\label{fig:dataset_single}
\end{figure}

\begin{figure}[H]
	\centering  
	\subfloat[Tufano's dataset]{
		\label{fig:dataset_density1}
		\includegraphics[width=0.24\textwidth]{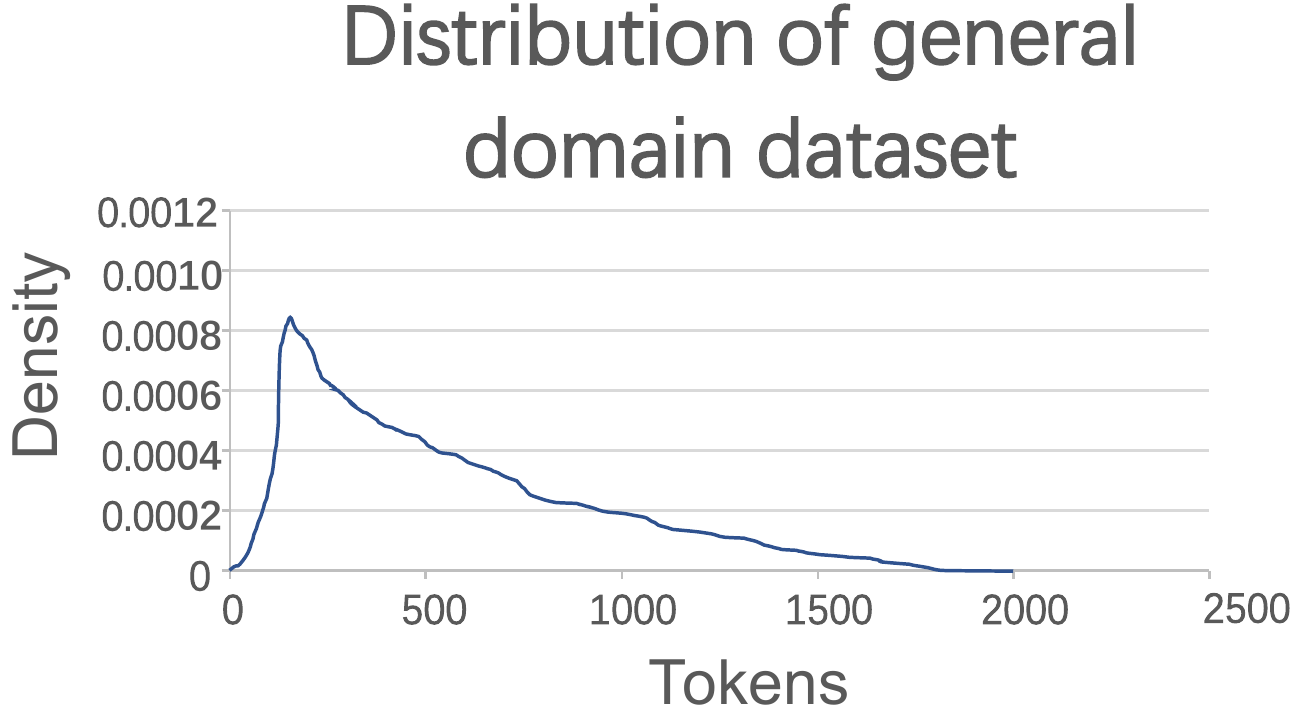}}
	\subfloat[Ponta's dataset]{
		\label{fig:dataset_density2}
		\includegraphics[width=0.24\textwidth]{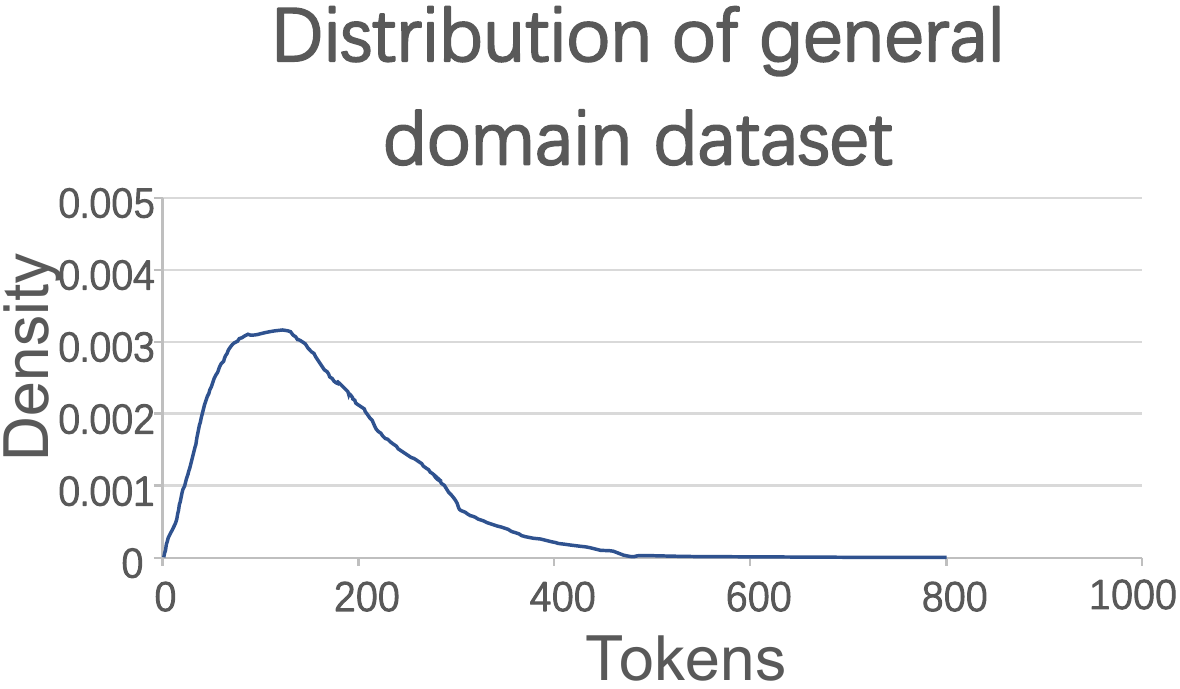}}
	\caption{Token distribution for each dataset}
	\label{fig:dataset_density}
\end{figure}

\begin{table}[H]	
	\renewcommand{\arraystretch}{1.3}
	\caption{Detailed information of $T_{tra}$}
	\label{tab:five_projects}
	\centering
	\scriptsize
	\begin{tabular}{|l|l|l|}
		\hline
		\textbf{Project Name} & \textbf{CVE Number} & \textbf{CWE Number} \\ \hline
		UAA & 37 & 10 \\ \hline
		Struts & 30 & 8 \\ \hline
		Spring-framework & 26 & 11 \\ \hline
		Lucene-solr & 14 & 2 \\ \hline
		Jenkins & 13 & 8 \\ \hline
	\end{tabular}
\end{table}


Figure~\ref{fig:dataset_single} shows the label distribution of each dataset. 
We can find that the frequency distribution of labels in the two datasets is very dissimilar.
Figure~\ref{fig:dataset_density} shows the token distribution of the vulnerable abstract context in each dataset.
It should be noted that the token length that is bigger than 2000 has been ignored in Tufano's dataset. 
The token length bigger than 800 has been ignored in Ponta's dataset.
We can find that most tokens in Tufano's dataset are distributed between 0 and 1500.
The majority of tokens in Ponta's dataset are distributed between 0 and 400.

In the third test set $T_{tra}$, we will try to use traditional evaluation approaches.
Five open-source projects which contain the largest number of CVEs (120 $ CPs $) will be selected as the test sets.
Fixing records on other projects will be used as training sets.
We will input suspicious files into the model of SeqTrans to generate patches.
We consider it a successful prediction if the predicted file passes the relevant test case and no new failures are introduced.
The detailed information of the test sets is shown in Table~\ref{tab:five_projects}.

In Table~\ref{tab:five_projects}, the first column shows the project name, including CloudFoundry User Account and Authentication Server (UAA), Apache Struts, Spring framework, Apache Solr and Jenkins.
Except for Apache Solr, every one of them has received more than 1K stars on Github.
Each of them has more than ten years of development history and has a stable maintenance team. 
We believe that their CVE fix records are relatively reliable and follow the specifications.
The second column shows the number of CVEs included in each project, and the third column shows the number of CWEs contained in each project.
It should be noted that nearly 5\% of the commit records were removed because they failed to pass compilation or the version was too old.
In addition, because these projects have long maintenance cycles and use different version control tools and development environments.
We manually configured all remaining project versions to ensure that each one would compile successfully and pass as many test cases as possible.

\subsubsection{\textbf{RQ1 Setup}:} 
The experimental part of RQ1 will be divided into three components, RQ1.1, RQ1.2 and RQ1.3.

Firstly, RQ1.1 will show and analyze the joint training and independent training results of the two datasets.
Since SeqTrans uses two datasets and a fine-tuning approach to overcome the problem of small samples, then independent and joint analyses for both datasets are necessary.
For the bug repair dataset of the general domain, we will train on $ G_{train} $ and validate on $ G_{val} $.
$ G_{val} $ is separated from the bug repair dataset, which is not contained in $ G_{train} $.
Likewise, we will separate the vulnerability dataset of specific domain to $ S_{train} $, $ S_{val} $ and $S_{test}$.
The $S_{test}$ will be utilized to validate the performance for both joint training and independent training.
Sequences in each set are mutually exclusive.
This experiment is designed to verify whether fine-tuning can help small samples overcome the problem of dataset size, learn from general domain tasks, and transfer it to specific domain tasks.

Secondly, RQ1.2 will compare SeqTrans with some state-of-the-art techniques such as Tufano \cite{tufano2018empirical, tufano2019learning} et al. and SequenceR \cite{chen2019sequencer}.
In order to avoid the effects of using pre-trained models, we will divide SeqTrans into SeqTrans\_full and SeqTrans\_single to refer to methods that use the pretrain model and the one that do not use the pretrain model.
SeqTrans\_full can be regarded as an enhancement of SeqTrans\_single as to alleviate the overfitting problem.
In the following sections, all SeqTrans that are not specified refer to SeqTrans\_full.

Tufano has investigated the feasibility of using neural machine translation for learning wild code. 
The disadvantage of his method is that only sentences with less than 100 tokens are analyzed.
SequenceR presents a novel end-to-end approach to program repair based on sequence-to-sequence learning.
It utilizes the copy mechanism to overcome the unlimited vocabulary problem. 
To the best of our knowledge, it achieves the best result reported on such a task.
However, the abstract data structure of this method retains too much useless context.
It does not use the normalization method either. 
We have also added the model that utilizes the same data structure but uses the seq2seq model.
Seq2seq model is an RNN encoder-decoder model widely used in the NMT domain, previous approaches such as SequenceR \cite{chen2019sequencer} and Tufano et al. \cite{tufano2018empirical} is also based on this model.
We have calculated the prediction accuracy for each technique.
Prediction accuracy will be calculated using 10-fold cross-validation for each technique.
Then we will calculate the number of correct predictions divided by the total number to calculate the accuracy.

Thirdly, RQ1.3 will apply SeqTrans on $T_{tra}$, the five projects selected from Ponta's dataset with the traditional evaluation approach.
Suspicious files will be input to the fine-tuned SeqTrans model to generate multiple patches.
The beam size is set to 10 but not 50 because it takes too long to compile and complete the test process.
The predicted and restored files will be sent back to the project to overwrite the source files. 
Then, we will recompile the whole project and run the test cases.
There is a vital problem here, how to define a vulnerability is successfully fixed? 
We will manually search and compare the parent commit of this CVE fix record.
If predicted files are compilable, all the diffs are semantically modified, and no new test failures are introduced, we consider it a correct fix.

Generated patches will be categorized into three types:
\begin{itemize}
	\item Compilable: The patch can pass the compiler.
	\item Plausible: The patch can pass the compiler and the test suite.
	\item Correct: The patch can pass the compiler and the test suite. It has also passed our manual checking.
\end{itemize}
These three types are inclusive relationships.
If the modified statement matches the changes in the commit, we consider it to be a correct patch.
If the modified statement does not match the changes in the commit, it will be manually determined if it affects the code logic. 
The plausible patches are manually checked by the first and the second author of this paper.
Both of them have more than five years of Java development experience. 

\subsubsection{\textbf{RQ2 Setup}:} In this part, we will discuss the impacts of the main factors that affect the performance of SeqTrans.

The process is shown as follows: Firstly, we will select a list of parameters that may affect the performance of our model.
Then we will change one parameter at one time and make the experiment in the same dataset.
We will utilize cross-validation ten times for each parameter and calculate the mean value as the final precision.
The final parameter selections of SeqTrans will produce the highest acceptance rates for the alternative configurations and data formats we tested.

\subsubsection{\textbf{RQ3 Setup}:} In this part, we will discuss the observations when we look deep inside the prediction results.
We only manually analyzed the prediction results generated by SeqTrans.
Other models are not considered.

We have calculated the prediction accuracy for each CWE and each category of code transformation.
We will look deep inside some well-predicted CWEs to explore why SeqTrans performs better on them.
We will also analyze why some CWEs have very poor prediction performance.

\subsection{Experimental Results}
\subsubsection{RQ1: How much effectiveness can SeqTrans provide for vulnerable code prediction?}
~\\
In RQ1, our goal is to analyze the performance of SeqTrans on the task of vulnerability fix.
As we have mentioned before, RQ1 will be divided into three components.
Firstly, we will analyze the joint training and independent training results of the two datasets in RQ1.1.
Table~\ref{tab:finetune} shows the prediction accuracy of models which were trained only on the general domain dataset (only on Tufano's dataset) or trained only on a specific domain dataset (only on Ponta's dataset) or trained jointly (fine-tuning strategy). 
The first column is the training approach of the three models.
The second column is the beam search size.
For example, in the situation of Beam=10, for each vulnerable sequence, we will generate ten prediction candidates.
If one of these ten candidates contains the correct prediction, the prediction accuracy is 1 otherwise it is 0.
The third column is the total prediction accuracy.
Recall that we use 10-fold cross-validation to calculate the accuracy of the model. 
If the predicted statement equals the statement in the test set, there is a correct prediction.

\begin{figure}[h]
	\centering
	\includegraphics[width=2.3in]{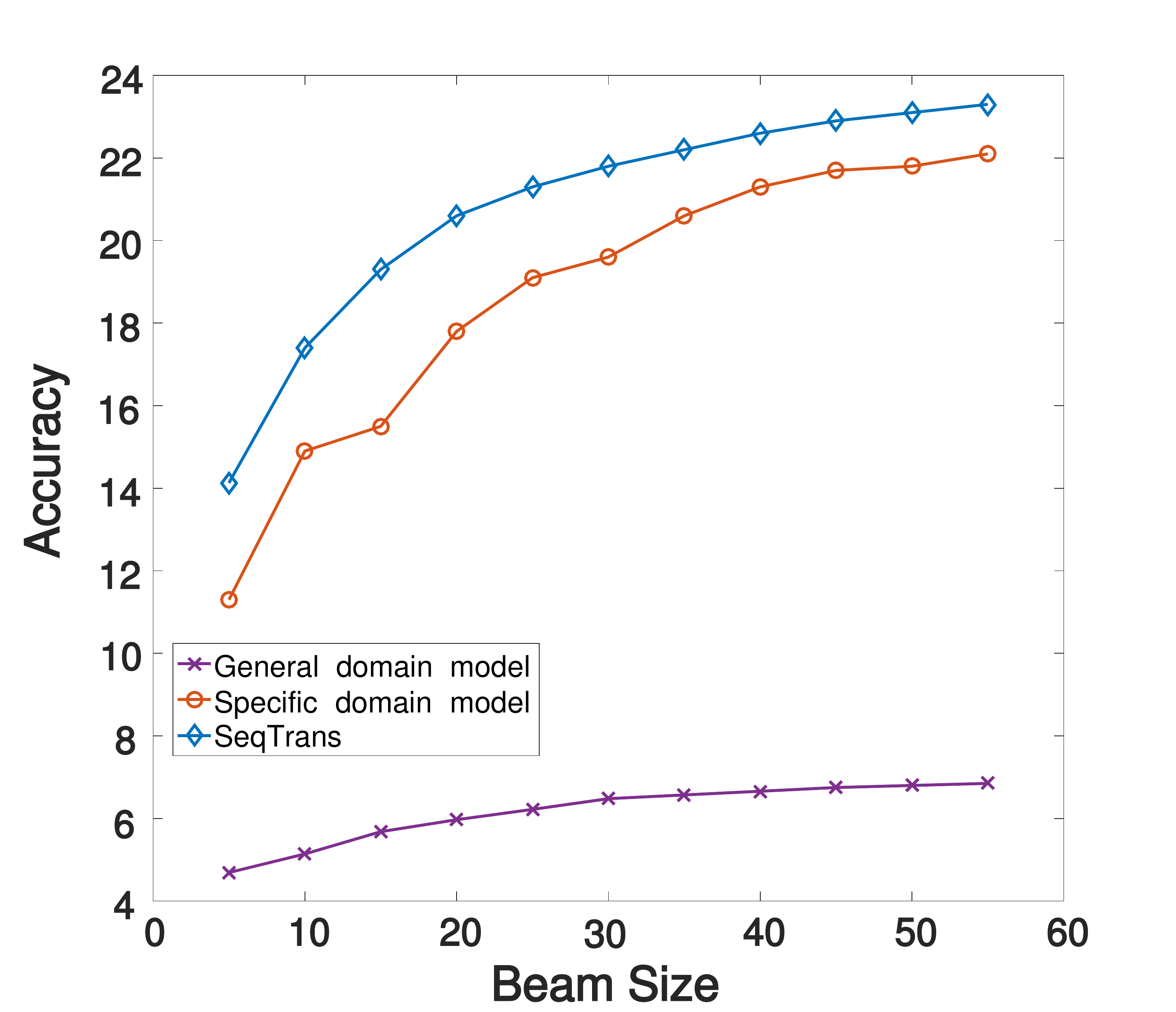}
	\caption{Performance of three training strategies}
	\label{fig:finetune}
\end{figure}

\noindent\textbf{RQ1.1:} From Table~\ref{tab:finetune}, we can observe that SeqTrans that use the fine-tuning strategy achieves the best performance of 14.1\% when Beam=1 and 23.3\% when Beam=50.
Next is the performance of 11.3\% when Beam=1 and 22.1\% when Beam=50 achieved by training on a specific domain dataset.
The worst prediction performance is using only data sets from the general domain, it can just achieve the accuracy of 4.7\% when Beam=1 and 6.9\% when Beam=50.
Detailed Beam search results are shown in Figure~\ref{fig:finetune} when beam size increases from 1 to 50.
The x-axis represents beam size and the y-axis represents the prediction accuracy.

Results show that using fine-tuning strategy to transfer knowledge from the general domain of bug repairing to the specific domain of vulnerability fixing improved the prediction performance of SeqTrans and achieved better performance than doing the training on two separate datasets.
Fine-tuning is helpful to alleviate and overcome the small data size problem.
In the following experiments, the fine-tuning strategy will become one of the default configurations in SeqTrans.

\noindent\textbf{RQ1.2:} Secondly, we will compare SeqTrans with some state-of-the-art techniques.
Table~\ref{tab:single_result} shows the accuracy results of single line prediction in five different NMT models including SeqTrans\_full, SeqTrans\_single, Seq2seq model, SequenceR, and the work of Tufano et al..
SeqTrans\_full, SeqTrans\_single refer to SeqTrans models that have been pre-trained and fine-tuned, and SeqTrans models that have been trained using only the Ponta's dataset.
For the Seq2seq model and transformer model, we use the same training set with def-use chains. As for the SequenceR~\cite{chen2019sequencer} and Tufano et al.~\cite{tufano2019learning}, we will strictly follow their original codes and data structures, repeat their preprocessing, training, and translating steps.

The reason why the total number in $ T_{cross} $ is inconsistent is that the data structure in different approaches is not the same.
SequenceR packages the entire class containing the buggy line, keeps the buggy method, all the instance variables, and only constructor's signature and non-buggy methods (stripping out the body).
Then it performs tokenization and truncation to create the abstract buggy context.
Because this abstract buggy context maintains too much context, even the whole buggy method and the constructor's signature in the class have the highest total number after deduplication.
Tufano et al. only construct the buggy pair that contains the buggy method and the corresponding fixed method.
However, they limit the whole sentence to 100 tokens and do not contain any statement outside of the method, so that this approach has the lowest total number after deduplication.
As introduced in Section~\ref{sec:ours}, our approach will maintain the buggy method with the vulnerable statement and any statement that has a data dependency on the vulnerable statement.
The total number of our approach is in the middle.

\begin{figure}[h]
	\centering
	\includegraphics[width=2.3in]{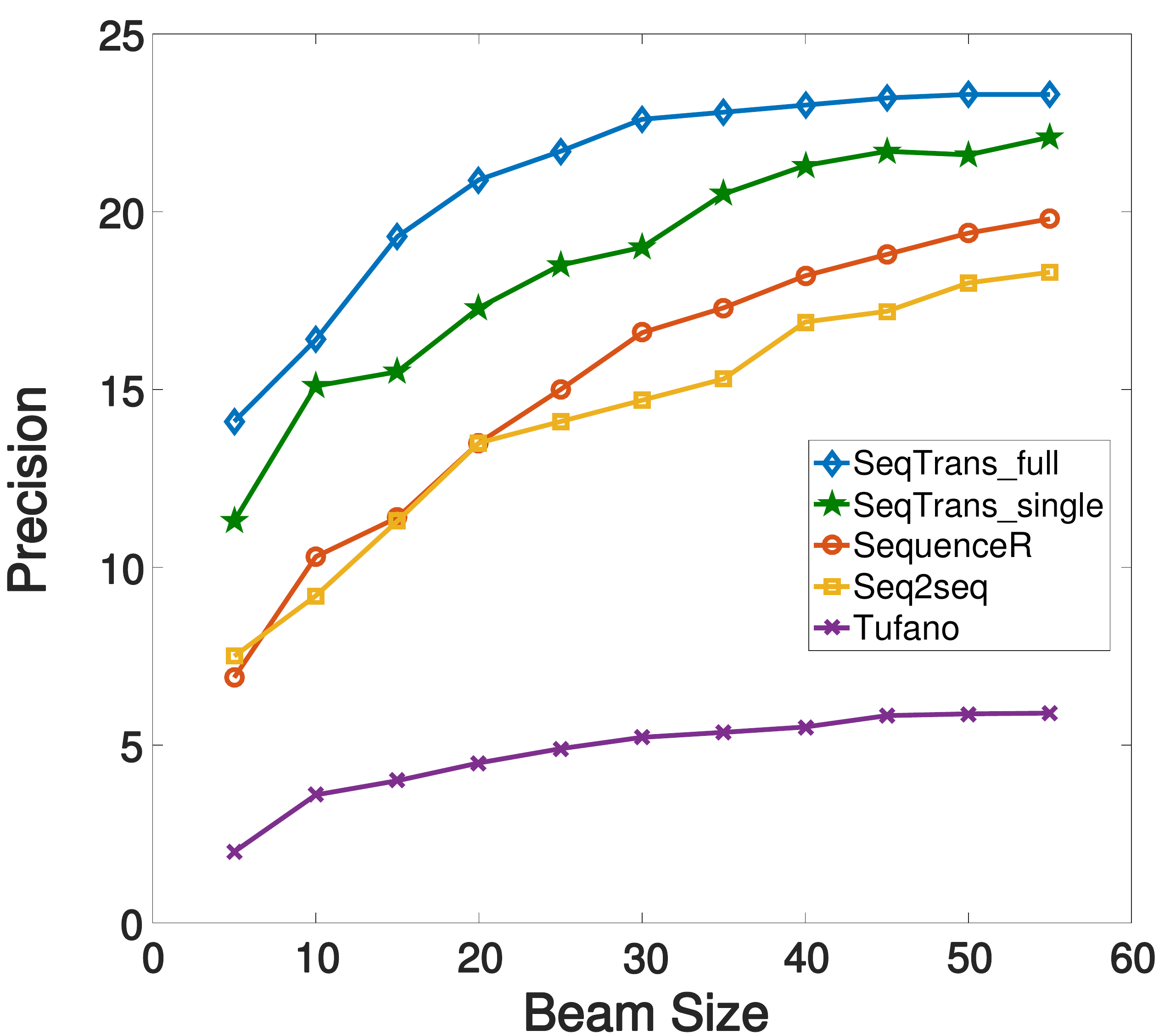}
	\caption{Performance of different techniques}
	\label{fig:single_noduplicate}
\end{figure}

In order to maintain a relatively fair training and testing environment, we introduce a second verification method.
As it has been explained previously, $ T_{cwe} $ provides an identical set of raw training, validation, and test dataset for each approach.
If one $ CP $ has been fully and correctly predicted, we regard it as a successful fix.
We have also tried to exploit the beam search to generate a list of predictions.
Figure~\ref{fig:single_noduplicate} shows the performance on $T_{cross}$ when beam size increases from 1 to 50.
The x-axis represents beam size and the y-axis represents the prediction accuracy.

From table~\ref{tab:single_result}, we see that our SeqTrans\_full performs the best and achieves an accuracy of 301/2130 (14.1\%) when Beam=1 on $ T_{cross} $, followed by SeqTrans\_single 338/2130 (11.3\%), Seq2seq 121/2130 (7.5\%), SequenceR 252/3661 (6.9\%) and Tufano et al. 37/883 (4.2\%).
On $ T_{cwe} $, SeqTrans\_full also reaches the best accuracy of 35/150(23.3\%) when Beam=1, followed by SeqTrans\_single 26/150 (17.3\%) SequenceR 24/150 (16.0\%), Seq2seq 20/150 (13.3\%) and Tufano et al. 5/150 (3.3\%).
The experimental results of $ T_{cross} $ and $ T_{cwe} $ are generally consistent.
We will do a more detailed case study in the RQ3.

To our surprise is that SequenceR is not as good as described.
It even performs worse than Seq2seq when Beam=1 on $ T_{cross} $. 
The difference between data structures can explain the poor performance of SequenceR.
SequenceR utilizes the buggy context, which contains the buggy line and the context around the buggy line in the same function.
Other variable and method declarations in the same class will also be retained.
However, this buggy context keeps many statements with no relationship with the buggy line.
The whole data structure is too long and contains numerous declaration statements unrelated to the buggy line, which performs poorly in our vulnerable public dataset.
Another disadvantage is that SequenceR only supports single-line prediction, but there are cases of statement deletions and additions included in the vulnerability fix.

In our SeqTrans, we only maintain the data dependencies before the vulnerable statement.
Meanwhile, we will normalize the data and replace variable names by $ "var1, var2....vark" $. 
The literals and numerical values will also be replaced by constants and maintained in a dictionary for future recovery.
The poor performance of Tufano et al. may be due to few data samples. 
We strictly follow their method and only select sequences with less than 100 tokens.
On the other hand, the fine-tuning method we use to learn from the general domain improves performance.
Another observation is that setting the beam size to 10 is sufficient in most cases.
Overall, SeqTrans leverages def-use chains and fine-tuning strategy to maintain data dependencies and overcome the minor data size issue, which can help the NMT model reach higher accuracy.

\noindent\textbf{RQ1.3:} Thirdly, we will use our SeqTrans to perform a traditional evaluation on five open source projects which contain the largest number of CVEs.
Table \ref{tab:projects_result} shows the results of these five projects.
The first column is the project name and the second column is the overall number. 
The third column is the compilable number, which means that at least one of the patches in this commit version is compilable.
The fourth column is the plausible number, it requires that the patch not only be compilable but also pass the test suite.
The fifth column is the number of correct patches, we will manually check the plausible patches to ensure these changes are semantically and functionally equivalent to the historical fixes.

\begin{figure}[h]
	\centering
	\includegraphics[width=\columnwidth]{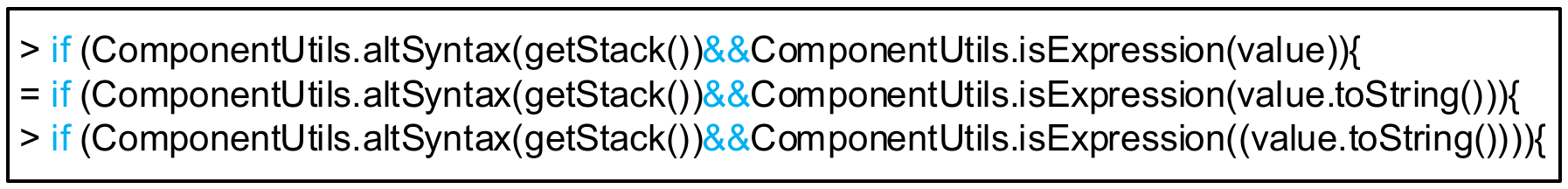}
	\caption{Case: fixing snippet of CVE-2016-0785 in Struts}
	\label{fig:RQ1.3_case}
\end{figure}

Results show that out of 120 vulnerabilities, SeqTrans generates at least one compilable patch for 98 vulnerabilities.
Some suspicious files cannot generate one compilable patch because some fixed records add or remove entire methods or rewrite the entire file.
For example, in \textit{SECURITY-499} of Jenkins, it rewrites two files and the associated test cases.
This case cannot be correctly fixed by our approach now.
SeqTrans also generates at least one plausible patch for 30 vulnerabilities.
This number is much smaller than the compilable number because many fixing histories not only modify source files but also change resource files such as the configuration files.
For another case, one fix may introduce new third-party packages.
This situation cannot be fixed by our approach now.
Finally, SeqTrans successfully generated at least one correct patch for 21 vulnerabilities.
We can see that nearly 18\% of the 120 vulnerabilities are fixed.
These patches have been manually checked to ensure they are semantically equivalent to the historical fixing records.
Figure~\ref{fig:RQ1.3_case} shows a fixing fragment of CVE-2016-0785, CWE-20 in Struts.
There is a pair of useless brackets in the prediction results of SeqTrans (the third line).
However, it does not influence the function of the statement.
In this case, we also treat it as a correct fix.

\begin{figure}[h]
	\centering
	\includegraphics[width=0.4\textwidth]{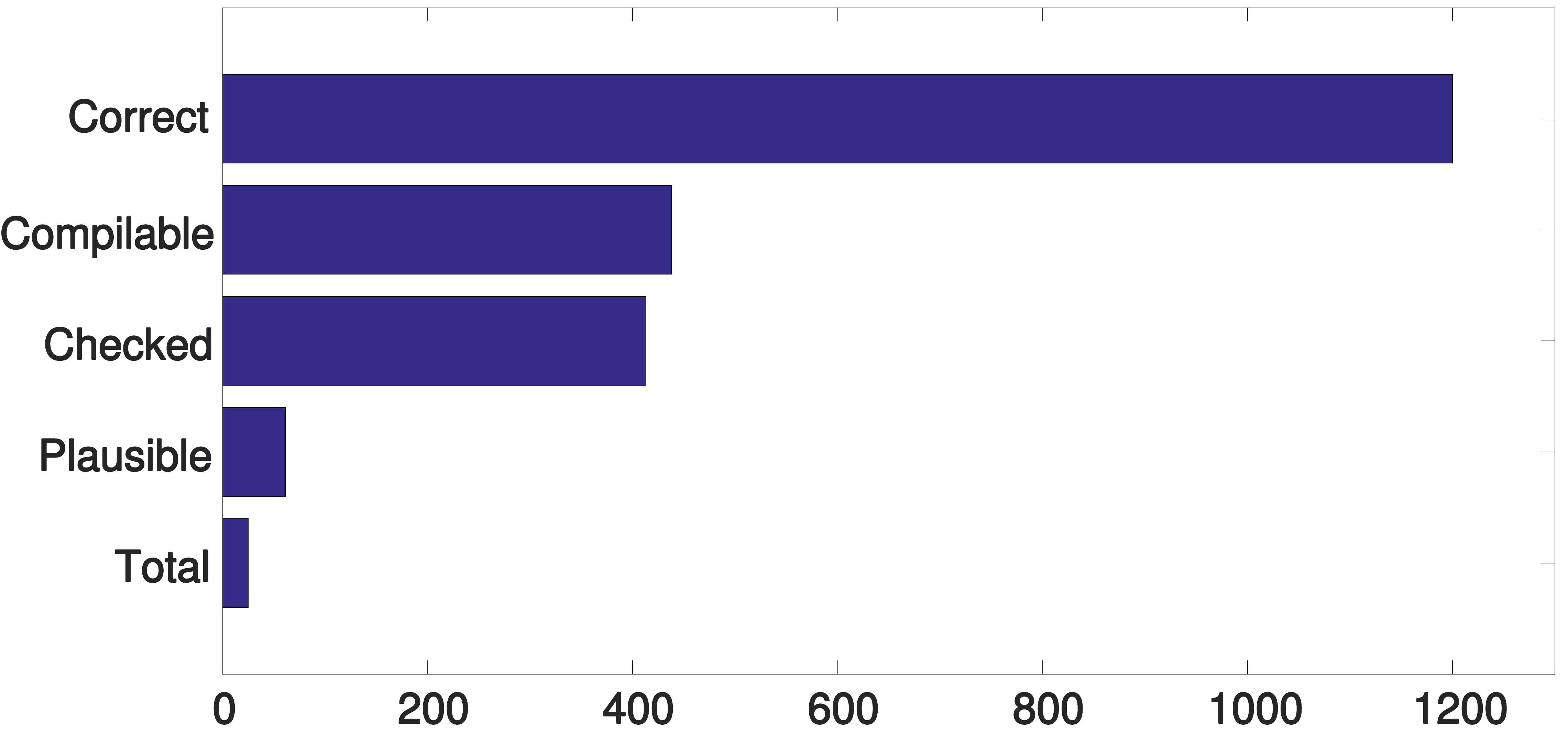}
	\caption{Statistics of SeqTrans generated patches for $ T_{tra} $}
	\label{fig:patches}
\end{figure}

Figure~\ref{fig:patches} shows a global statistic for $ T_{tra} $. 
In the figure, we add a checked tag to analyze the effectiveness of FindBugs, which means the patch that has passed the static analysis check.
For a total of 1200 generated patches, 438 patches can be compiled.
Then, after the checking of FindBugs, 413 patches are survived.
In these patches, 49 of them are plausible, and finally, 25 patches are validated to be correct, which means they have passed the relevant test cases and are semantically equivalent to historical fix records.
Here we give some observations from Figure~\ref{fig:patches}.
The compiler filters out most of the 787 invalid patches filtered by the compiler and the checker.
The checker only filtered out 25 patches.
FindBugs actually reports more numbers than this, but most of them are not associated with the vulnerable statements.
The total plausible number in Figure~\ref{fig:patches} is larger than Tabel~\ref{tab:projects_result}, which means there is more than one plausible patch for one CVE.
This situation can heavily rely on the quality of the related test cases~\cite{martinez2017automatic}.
This gap will be reduced if the developer commits the relevant test set changes together with the commit promptly.
This result makes us consider whether removing the checking part to reduce the overhead is good.
We will explore more options in our future work.

In general, the current functionality of SeqTrans is suitable as assistance to developers for program repair. 
There is still a long way to separate from the developers and independently do accurate automatic program fixes.

\begin{table}[h]
	\renewcommand{\arraystretch}{1.6}
	\caption{Prediction results in three training strategies}
	\label{tab:finetune}
	\centering
	\scriptsize
	\begin{tabular}{|l|l|l|}
		\hline
		\textbf{Approach} & \textbf{Beam} & \textbf{Accuracy} \\ \hline
		\multirow{3}{*}{Only on general domain $ G_{train} $} & 1 & 100/2130(4.7\%) \\ \cline{2-3} 
		& 10 & 121/2130(5.7\%) \\ \cline{2-3} 
		& 50 & 146/2130(6.9\%) \\ \hline
		\multirow{3}{*}{Only on specific domain $ S_{train} $} & 1 & 242/2130(11.3\%) \\ \cline{2-3} 
		& 10 & 338/2130(15.5\%) \\ \cline{2-3} 
		& 50 & 473/2130(22.1\%) \\ \hline
		\multirow{3}{*}{Joint training on $ G_{train} $ and $ S_{train} $} & 1 & 301/2130(14.1\%) \\ \cline{2-3} 
		& 10 & 411/2130(19.3\%) \\ \cline{2-3} 
		& 50 & 497/2130(23.3\%) \\ \hline
	\end{tabular}
\end{table}

\begin{table}[h]	
	\renewcommand{\arraystretch}{1.4}
	\newcommand{\tabincell}[2]{\begin{tabular}{@{}#1@{}}#2\end{tabular}}
	\caption{Effectiveness on the five selected projects}
	\label{tab:projects_result}
	\centering
	\scriptsize
	\begin{tabular}{|l|l|l|l|l|}
		\hline
		\textbf{Project Name} & \textbf{Total} & \textbf{Compilable} & \textbf{Plausible} & \textbf{Correct} \\ \hline
		UAA & 37 & 31 & 9 & 6 \\ \hline
		Struts & 30 & 25 & 10 & 7 \\ \hline
		Spring-framework & 26 & 21 & 6 & 4 \\ \hline
		Lucene-solr & 14 & 11 & 3 & 2 \\ \hline
		Jenkins & 13 & 9 & 2 & 2 \\ \hline
	\end{tabular}
\end{table}

\begin{table}[h]
	\renewcommand{\arraystretch}{1.6}
	\newcommand{\tabincell}[2]{\begin{tabular}{@{}#1@{}}#2\end{tabular}}
	\caption{Performance of different techniques}
	\label{tab:single_result}
	\centering
	\scriptsize
	\begin{tabular}{|l|l|l|l|}
		\hline
		\multirow{2}{*}{\textbf{Approach}} & \multirow{2}{*}{\textbf{Beam}} & \multicolumn{2}{c|}{\textbf{Accuracy}} \\ \cline{3-4} 
		&  & \multicolumn{1}{c|}{$ T_{cross} $} & \multicolumn{1}{c|}{$ T_{cwe} $} \\ \hline
		\multirow{3}{*}{SeqTrans\_full} & 1 & 301/2130(14.1\%) & 35/150(23.3\%) \\ \cline{2-4} 
		& 10 & 411/2130(19.3\%) & 38/150(25.3\%) \\ \cline{2-4} 
		& 50 & 497/2130(23.3\%) & 38/150(25.3\%) \\ \hline
		\multirow{3}{*}{SeqTrans\_single} & 1 & 242/2130(11.3\%) & 26/150(17.3\%) \\ \cline{2-4} 
		& 10 & 338/2130(15.5\%) & 31/150(20.7\%) \\ \cline{2-4} 
		& 50 & 473/2130(22.1\%) & 31/150(20.7\%) \\ \hline
		\multirow{3}{*}{SequenceR} & 1 & 252/3660(6.9\%) & 24/150(16.0\%) \\ \cline{2-4} 
		& 10 & 418/3660(11.4\%) & 26/150(17.3\%) \\ \cline{2-4} 
		& 50 & 725/3660(19.8\%) & 27/150(18.0\%) \\ \hline
		\multirow{3}{*}{Seq2seq} & 1 & 121/2130(7.5\%) & 20/150(13.3\%) \\ \cline{2-4} 
		& 10 & 242/2130(11.3\%) & 23/150(15.3\%) \\ \cline{2-4} 
		& 50 & 390/2130(18.3\%) & 23/150(15.3\%) \\ \hline
		\multirow{3}{*}{Tufano et al.} & 1 & 37/883(4.2\%) & 5/150(3.3\%) \\ \cline{2-4} 
		& 10 & 59/883(6.7\%) & 7/150(4.6\%) \\ \cline{2-4} 
		& 50 & 63/883(7.1\%) & 7/150(4.6\%) \\ \hline
	\end{tabular}
\end{table}

\vspace{1em}
\noindent\fbox{
	\parbox{0.95\linewidth}{
		\textbf{Answer to RQ1}:
		In summary, NMT models are able to learn meaningful code changes from historical code repair records and generate predicted patch to assist developers with code repairs.
		Our approach SeqTrans based on a transformer model outperforms other NMT models on the task of vulnerability fixing.
		Even it outperforms the state-of-the-art approach SequenceR in our public vulnerability fix dataset.
}}

\subsubsection{RQ2: What are the characteristics of the ML model used that can impact the performance of SeqTrans?}

\begin{table}[t]
	\renewcommand{\arraystretch}{1.6}
	\caption{Factor impact analysis with selected parameters}
	\label{tab:factor_result}
	\centering
	\scriptsize
	\begin{tabular}{|c|l|c|c|}
		\hline
		\multicolumn{1}{|l|}{\textbf{Group}} & \textbf{Description} & \multicolumn{1}{l|}{\textbf{Precision}} & \multicolumn{1}{l|}{\textbf{Impact}} \\ \hline
		- & Default SeqTrans model & 23.3\% & - \\ \hline
		1 & Word Size (256 vs 512) & 22.4\% & -4\% \\ \hline
		& Word Size (512 vs 1024) & 22.1\% & -5\% \\ \hline
		2 & Training steps (30K vs 100K) & 23.5\% & 1\% \\ \hline
		3 & Layers (5 vs 6) & 21.9\% & -6\% \\ \hline
		& Layers (6 vs 7) & 22.4\% & -4\% \\ \hline
		4 & Batch Size (2048 vs 4096) & 22.6\% & -3\% \\ \hline
		5 & Hidden State Size (256 vs 512) & 22.8\% & -2\%  \\ \hline
		6 & Without Def-use Chains & 20.9\% & -10\% \\ \hline
		7 & Without Code Normalization & 21.9\% & -6\% \\ \hline
		8 & Without BPE & 23.3\% & 0\% \\ \hline
		9 & Without Mixed Fine-tuning & 22.1\% & -5\% \\ \hline
		10 & Without Fine-tuning Strategy & 20.2\% & -13\% \\ \hline		
	\end{tabular}
\end{table}

\begin{figure}[!t]
	\centering
	\subfloat[Layers]{
		\label{fig:layers}
		\includegraphics[width=0.23\textwidth]{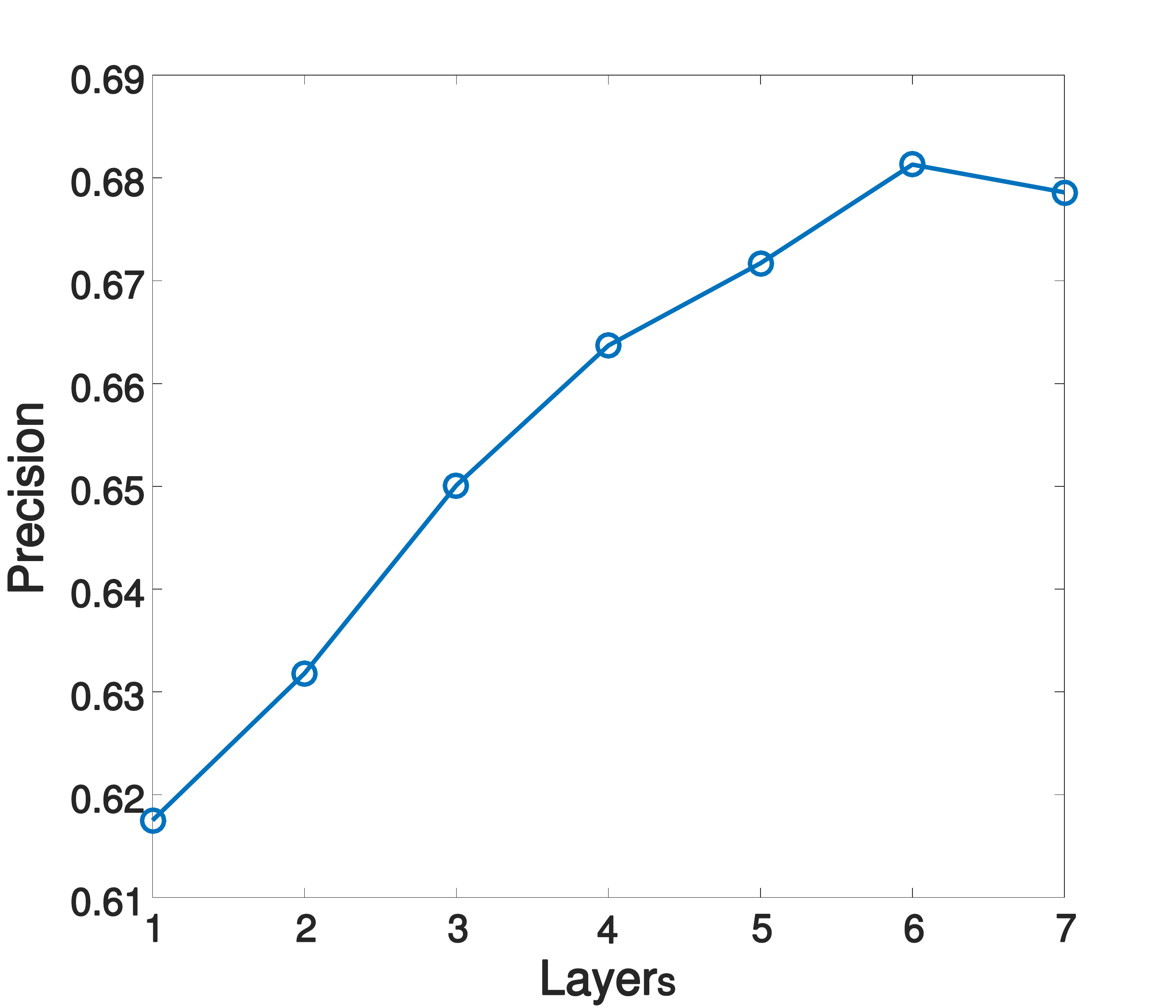}}
	\subfloat[Training Steps]{
		\label{fig:steps}
		\includegraphics[width=0.25\textwidth]{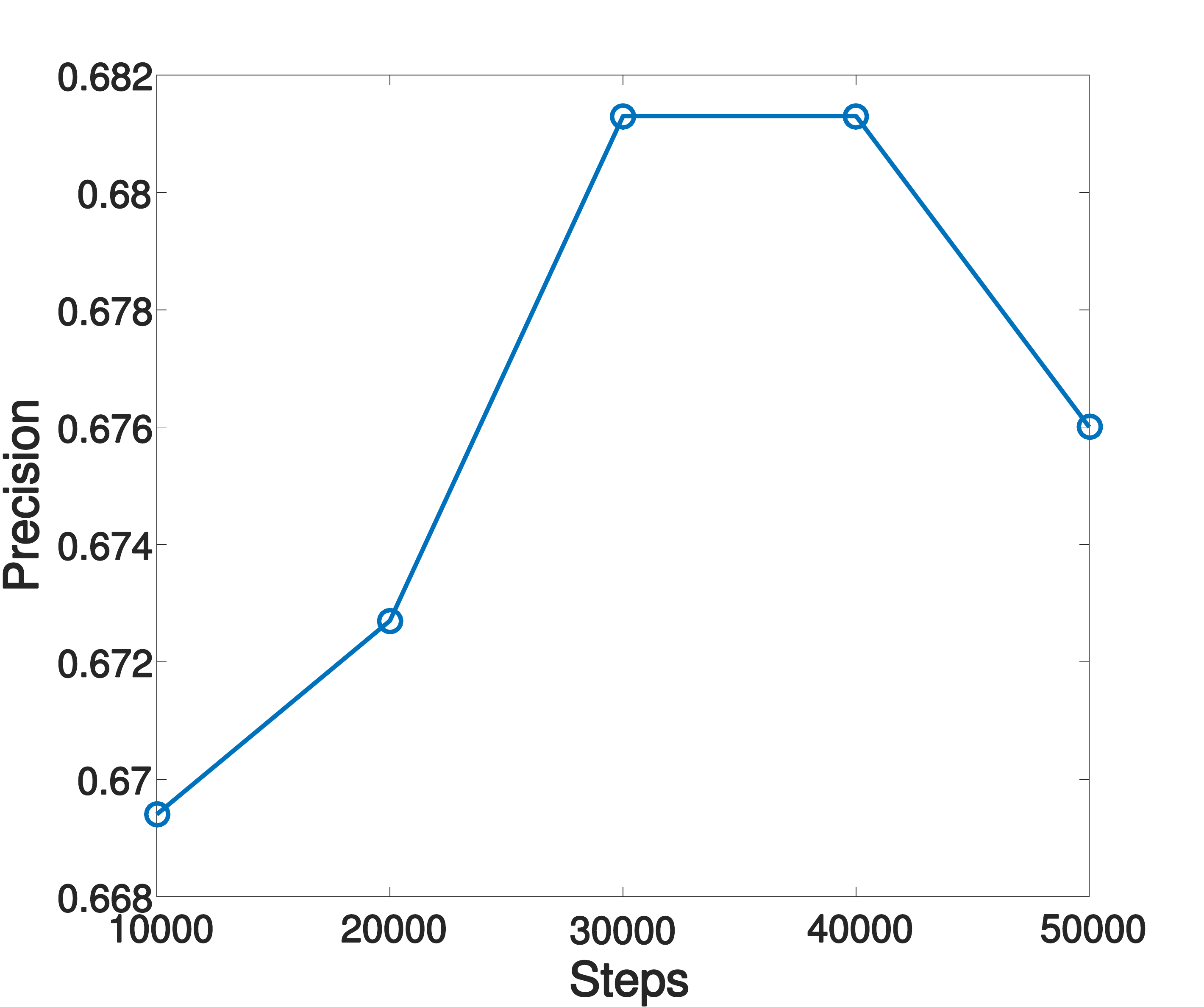}}
	\caption{Factor analysis with selected parameters}
	\label{fig:parameters}
\end{figure}

In RQ2, we will discuss some of the data formats and configuration exploration processes that we have tried to get a default SeqTrans model eventually.
Table~\ref{tab:factor_result} and Figure~\ref{fig:parameters} shows an ablation study for SeqTrans.
From Table~\ref{tab:factor_result}, we can see the prediction result of our default SeqTrans against the results of single changes on the model. We will explain them one by one.
These ablation results will help future researchers understand which configurations are most likely to improve their own models.
Due to the random nature of the learning process, we will use the 10-fold cross-validation on $ T_{cross} $ to train each control group 10 times and take the mean value as the final result.
The first row is the performance of the default SeqTrans model as a reference.

Group 1 in the second and third rows explored the effect of word size on the performance of our model.
Results show that both the smaller and larger word sizes perform worse than the configuration we choose.
We think the reason is that Smaller word sizes may lead to transitional compression of features and loss of some valid information. 
Larger word sizes may not be appropriate for the size of our dataset.

In Group 2 and Figure~\ref{fig:steps} we have discussed whether more training steps would significantly improve performance.
The result indicates that the performance difference between 30K and 100K training steps is very small.
The growth in predicted performance begins to converge after 30k training steps.
We do not consider it worthwhile due to the large time overhead of 100K training steps.
It is worth noting that the training step here refers to the step used when fine-tuning the dataset of vulnerability fixing tasks in the special domain, and the general domain model is consistent.

Group 3 in the fifth and sixth rows and Figure~\ref{fig:layers} are the test of model layers, we have tried different features and the conclusion is that 6 layers are a suitable choice.
It is worth noting that we need to ensure that the encoder and decoder parts of the transformer model have the same number of layers, so we use the same number of layers on both the encoder and decoder.
Results show that prediction performance rises with the number of layers until it reaches 6.
The performance of layer 7 is not better than 6, so we decide on 6 as the parameter.
Group 4 and Group 5 are the test of different batch sizes and hidden state sizes.
The experimental results show a similar conclusion: decreasing the size leads to decreased performance.

In group 6, 7 and 8, we will discuss the impact of data structure and processing on performance.
The result shows a 10\% improvement in model performance when comparing our data structure to the original single vulnerable line.
Normalization in data preprocessing will lead to a 6\% increase in performance.
An interesting phenomenon is that whether BPE is enabled or not has only a minimal performance impact.
We think the main purpose of BPE is to compress the data and solve the problem of unregistered words.
Our vocabulary size is able to cover the majority of words.
However, when we prepare the first general model, not using BPE to compress the sequences will cause a huge vocabulary size and lead to the overflow of GPU memory.

Group 9 is designed to explore whether mixing some general domain training data into the small specific domain dataset can alleviate the problem of catastrophic forgetting.
We tried to mix in the same number of randomly selected $ G_{train} $ training data as $ S_{train} $ and compare the results with the original $S_{train} $ experiments.
The result shows that without mixing the prediction performance indeed causes a degradation of the performance.
The last Group 10 is the performance change before and after using the fine-tuning strategy as explained in the previous experiments.
SeqTrans achieves a 13\% performance improvement, indicating that the fine-tuning strategy is very beneficial for training small-scale data and helps us migrate knowledge from similar domains.

\vspace{1em}
\noindent\fbox{
	\parbox{0.95\linewidth}{
		\textbf{Answer to RQ2}:
		The ablation study results demonstrate that parameter selections for the SeqTrans produce the highest acceptance rates for the configurations we tested.
		These ablation results will help future researchers understand which configurations are most likely to improve their own models.
}}

\subsubsection{RQ3: How does SeqTrans perform in predicting specific types of CWEs?}
~\\
We now look at what types of vulnerabilities fix our model can well identify and generate predictions.
The purpose of this experiment is to verify whether SeqTrans has better performance for a specific type of CWE. 
For example, the CWEs have a high number of repair cases in the dataset or the CWEs are uniformly distributed in the data set by time series.
Table~\ref{tab:CWE_result} shows the prediction accuracy of each CWE in $ T_{cross} $ and $ T_{cwe} $ when Beam=50.
The Common Weakness Enumeration (CWE) is a category system for software weaknesses and vulnerabilities. 
Every CWE contains a list of CVEs.
Because there are too many kinds of CWE, we only list the top 20 with the highest accuracy in the table, which contains the vast majority of correct predictions.
It should be mentioned that the total result may be higher than the results in Table~\ref{tab:single_result}.
The reason is that some CVE may belong to multiple kinds of CWE.
It will be counted multiple times when counting the number of CWEs.

Then we will explain Table~\ref{tab:CWE_result}.
As for $ T_{cross} $, the highest one is CWE-444, which achieves the accuracy of 60\%.
If only the highest number of predictions is considered, it is CWE-502, which contains 311 correct predictions.
As for $ T_{cwe} $, the highest one is CWE-306 and it achieves a surprising prediction performance of 100\%.
If only the highest number of predictions is considered, it is CWE-22, which contains ten correct predictions.
Detailed results are given in Table \ref{tab:CWE_result}.
\textit{CWE No.} indicates the CWE number.
The first column of \textit{Accu} is the right prediction number and the total prediction number.
The second column of \textit{Accu} is prediction accuracy.
We can find that most of the TOP CWE predictions in the two test sets are the same.
CWEs with large differences will be labeled.
CWEs in $ T_{cwe} $ contain less CWE categories than $ T_{cross} $, which may have contributed to the greater concentration of top CWE.
In the following, we will compare the difference between these two test sets and make a detailed analysis of why the model performs well on certain specific CWEs.
They perform differently or even achieve zero accuracies in one dataset.
First of all, it must be stated that the reason why these CWEs marked blue are not present on the right side is that they are not included in $ T_{cwe} $.
These will not be the focus of our attention.

%

\begin{table}[]
	\renewcommand{\arraystretch}{1.3}
	\caption{Prediction results in the data set}
	\label{tab:CWE_result}
	\centering
	\scriptsize
	\begin{tabular}{|l|l|c|l|l|c|}
		\hline
		\multicolumn{3}{|c|}{\textbf{$ T_{cross} $}} & \multicolumn{3}{c|}{\textbf{$ T_{cwe} $}} \\ \hline
		\textbf{CWE No.} & \multicolumn{2}{l|}{\textbf{Accu}} & \textbf{CWE No.} & \multicolumn{2}{l|}{\textbf{Accu}} \\ \hline
		{\color{blue} CWE-444} & 3/5 & 0.60 & CWE-306 & 1/1 & 1.00 \\ \hline
		CWE-287 & 45/84 & 0.54 & CWE-287 & 2/3 & 0.67 \\ \hline
		CWE-306 & 1/2 & 0.50 & CWE-20 & 8/14 & 0.57 \\ \hline
		{\color{red} CWE-362} & 5/11 & 0.45 & CWE-522 & 2/4 & 0.50 \\ \hline
		CWE-22 & 13/30 & 0.43 & CWE-22 & 10/21 & 0.48 \\ \hline
		{\color{blue} CWE-361} & 3/7 & 0.43 & CWE-295 & 1/3 & 0.33 \\ \hline
		CWE-863 & 7/17 & 0.41 & CWE-269 & 1/3 & 0.33 \\ \hline
		{\color{blue} CWE-284} & 3/8 & 0.38 & CWE-863 & 3/10 & 0.30 \\ \hline
		CWE-522 & 24/67 & 0.36 & CWE-502 & 5/12 & 0.42 \\ \hline
		CWE-20 & 31/97 & 0.32 & CWE-611 & 3/13 & 0.23 \\ \hline
		{\color{red} CWE-502} & 311/1013 & 0.31 & CWE-200 & 2/11 & 0.18 \\ \hline
		{\color{red} CWE-78} & 7/23 & 0.30 & CWE-noinfo & 2/13 & 0.15 \\ \hline
		{\color{red} CWE-74} & 4/14 & 0.29 & CWE-78 & 0/5 & 0 \\ \hline
		CWE-310 & 41/147 & 0.28 & CWE-35 & 0/3 & 0 \\ \hline
		CWE-269 & 8/29 & 0.28 & CWE-601 & 0/2 & 0 \\ \hline
		{\color{blue} CWE-264} & 14/60 & 0.23 & CWE-74 & 0/2 & 0 \\ \hline
		{\color{blue} CWE-611} & 1/52 & 0.21 & CWE-362 & \multicolumn{1}{c|}{0/1} & 0 \\ \hline
		CWE-noinfo & 7/54 & 0.13 & CWE-521 & \multicolumn{1}{c|}{0/1} & 0 \\ \hline
		{\color{blue} CWE-200} & 3/28 & 0.11 & CWE-50 & \multicolumn{1}{c|}{0/1} & 0 \\ \hline
		{\color{blue} CWE-19} & 5/56 & 0.09 & CWE-89 & \multicolumn{1}{c|}{0/1} & 0 \\ \hline
		All & 563/2130 & 26.4\% & All & \multicolumn{1}{c|}{40/150} & 26.7\% \\ \hline
	\end{tabular}
\end{table}

\begin{figure}[htbp]
	\centering
	\includegraphics[width=\columnwidth]{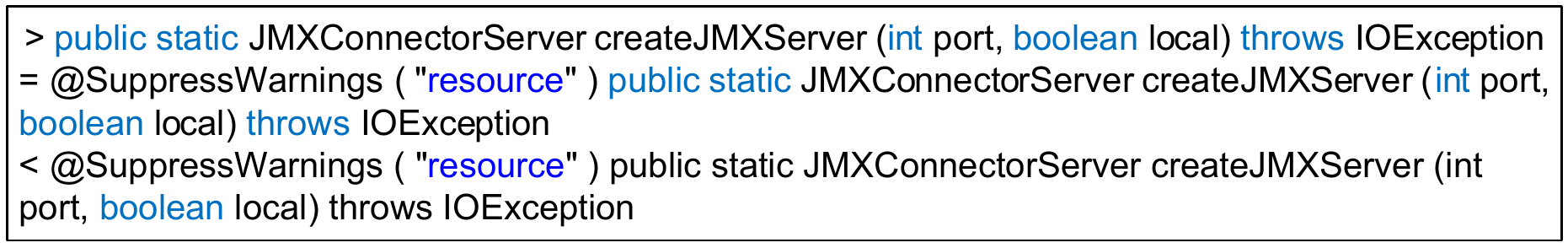} \\
	\includegraphics[width=\columnwidth]{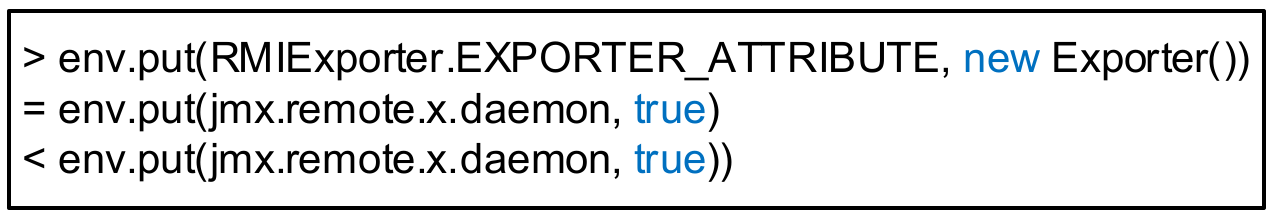}
	\caption{Case: right prediction of CWE-306}
	\label{fig:case-306}
\end{figure}

\textbf{Case Study: CWE-306:} CWE-306 means "Missing Authentication for Critical Function".
It is special because it has a very small sample but makes a correct prediction.
The software does not perform any authentication for functionality requiring a provable user identity or consuming significant resources.
This commit contains two code changes as shown in Figure~\ref{fig:case-306}.
The first one (second line) is to add the annotation "@SuppressWarnings ( "resource" )" before the method declaration.
The second one is to modify two parameters in the put method.

These two modifications have been correctly captured and predicted by SeqTrans.
The other two incorrect predictions belong to variable definition changes, the model does not make the correct prediction.

%

\textbf{Case Study: CWE-362:} CWE-362 means "Concurrent Execution using Shared Resource with Improper Synchronization".
The program contains a code sequence that can run concurrently with other code, and the code sequence requires temporary, exclusive access to a shared resource, but a timing window exists in which the shared resource can be modified by another code sequence that is operating concurrently.
It contains a list of condition operator changes and parallelism-related modifications.

\begin{figure}[htbp]
	\centering
	\includegraphics[width=\columnwidth]{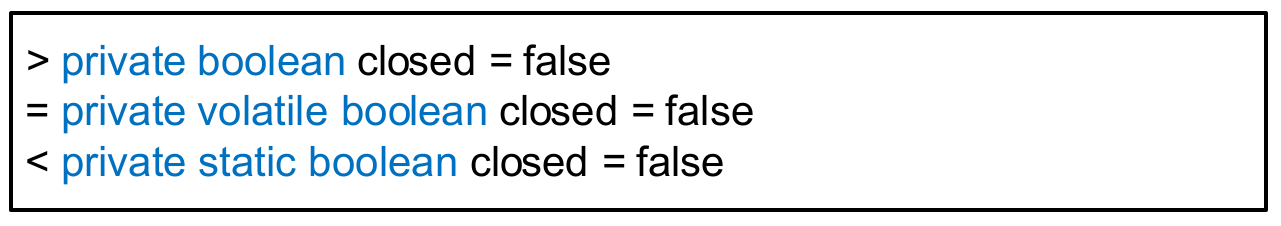} \\
	\includegraphics[width=\columnwidth]{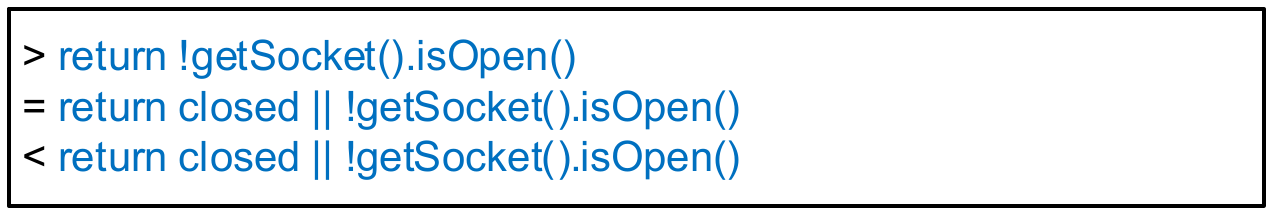}
	\caption{Case: wrong prediction of CWE-362}
	\label{fig:case-362}
\end{figure}

In Figure~\ref{fig:case-362}, developers added one keyword and changed the return condition.
The condition modification of the statement has been correctly predicted by SeqTrans.
However, the addition of the volatile keyword was not successfully predicted by $ T_{cwe} $'s model.
We think the reason is that $ T_{cross} $'s model learns from other records about adding the static keyword.

\textbf{Case Study: CWE-502:} CWE-502 means "Deserialization of Untrusted Data".
The application deserializes untrusted data without sufficiently verifying that the resulting data will be valid.
CWE-502 related code transformations account for half of the entire training set. 
It contains large numbers of repetitive code transformations, such as deleting one throw exception, adding a return statement, and changing parameter orders.
We will list some typical code changes that are well captured and handled by SeqTrans. 

\begin{figure}[H]
	\centering
	\includegraphics[width=\columnwidth]{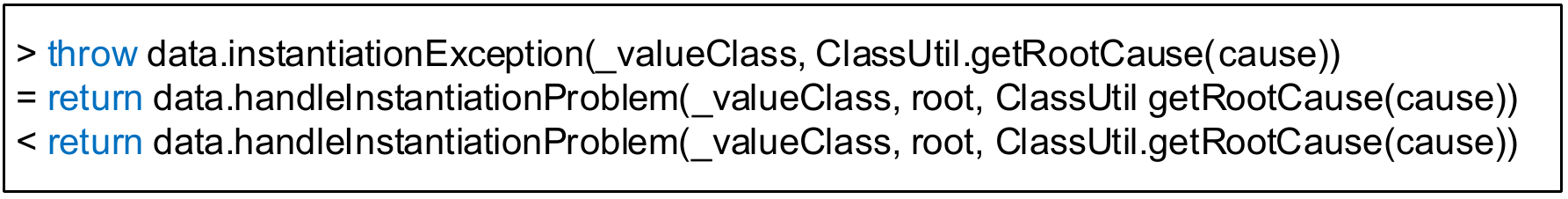}
	\caption{Case: right prediction of CWE-502}
	\label{fig:case-502-1}
	\vspace{-0.3cm}
\end{figure}

In Figure \ref{fig:case-502-1}, developers delete the throw keyword and add a return keyword to transfer the instantiation problem.
In addition, a new parameter was inserted into the second position.
This code transformation can be well captured by SeqTrans.

\begin{figure}[H]
	\centering
	\includegraphics[width=1.8in]{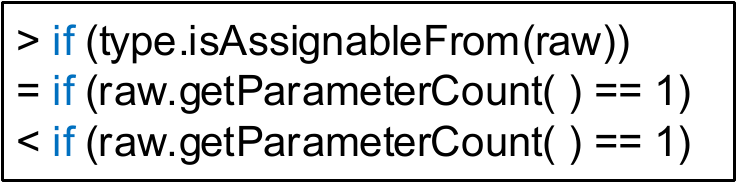}
	\caption{Case: right prediction of CWE-502}
	\label{fig:case-502-2}
\end{figure}

In Figure \ref{fig:case-502-2}, developers firstly change the target of the method call. 
Then, replace the method call from "isAssignableFrom" to "getParameterCount".
Finally, the conditional expression "== 1" is added.
This code transformation contains three single code transformations but is also well captured by SeqTrans.
In general, our tool SeqTrans performs stable and outstandingly for vulnerability fixes like CWE-502 that contain a lot of repetitive code transformations.

\textbf{Case Study: CWE-78 and CWE-74:}
These two CWEs face the same problem and we will explain them together.
CWE-78 means "Improper Neutralization of Special Elements used in an OS Command".
The software constructs all or part of an OS command using externally-influenced input from an upstream component, but it does not neutralize or incorrectly neutralize special elements that could modify the intended OS command when sent to a downstream component.
CWE-74 means "Improper Neutralization of Special Elements in Output Used by a Downstream Component".
The software constructs all or part of a command, data structure, or record using externally-influenced input from an upstream component, but it does not neutralize or incorrectly neutralize special elements that could modify how it is parsed or interpreted when it is sent to a downstream component.
We give the following explanation for the 0\% accuracy of these two CWEs: $ T_{cwe} $ does not contain any of them in the training set.
All of them are included in the test set.
We believe that this situation is the cause of the low accuracy rate.

The conclusion reached is that, for some CWEs that contain duplicate vulnerability fixes or can be learned from historical repair records, our SeqTrans performs very well.
Another hypothesis is that training a general model to fix vulnerabilities automatically is too ambitious to cover all cases. 
If we can focus on specific types of CWEs, the NMT model can make a very promising result to help developers.
~\\
~\\
\noindent\fbox{
	\parbox{0.95\linewidth}{
		\textbf{Answer to RQ3. Finding 1:} SeqTrans performs well in predicting specific kinds of vulnerability fixes like CWE-287 and CWE-362. It also performs well on a timing test set that simulates learning historical modification records.
		The prediction range will become wider and wider as the historical repair records increases.
}}

~\\
~\\
On the other hand, to deeply analyze these specific CWEs, we derived Table~\ref{tab:trans_result} that shows the classification of code transformations by manually analyzing prediction results and source codes.
We have made a change type classification for each code change not only the correct prediction but also the wrong prediction.
We only consider the prediction results strictly consistent with the true modifications as correct predictions.
So the actual accuracy should be higher than the strict matching calculation method we used.
The first column is the type name of code transformations.
We roughly divided the code transformation types into 17 categories.
It is worth noting that some single predictions can include multiple types of code changes, they are classified into different code change types.
For this reason, the sum of the classified changes is not equaled to the number in Table \ref{tab:CWE_result}.
Detailed definitions are shown in the following:

\begin{itemize}
	\item Change Parameter: Add, delete the parameter or change the parameter order.
	\item Change Throw Exception: Add, delete or replace the block of throw exception, add or delete the exception keywords in the method declaration.
	\item Change Variable Definition: Change variable type or value.
	\item Change Method Call: Add, delete a method call or replace a method call by another.
	\item Change Target: Maintain the same method call but change the target of the method call.
	\item Change String: Add, delete or replace the string.
	\item Change Method Declaration: Add, delete or replace method name and the qualifier.
	\item Change Class Declaration: Modify the declaration of a class.
	\item Change if Condition: Add, delete or replace operands and operators in the if condition.
	\item Change Switch Block: Add, delete or replace the "case" statement.
	\item Change Loop Condition: Modify the loop condition.
	\item Change Return Statement: Change return type or value, add or delete "return" keyword.
	\item Change Keywords "this/super": add or delete these keywords.
	\item Change Try Block: Put statements into the try block.
	\item Change Catch Exception: Add, delete or replace the block of catch exception.
	\item Refactoring: Rewrite the code without changing functionality. 
	\item Other: Other transformations which are hard to be categorized or occur infrequently.
\end{itemize}

\begin{table}[t]
	\renewcommand{\arraystretch}{1.3}
	\caption{Types of code transformation learned by SeqTrans}
	\label{tab:trans_result}
	\centering
	\scriptsize
	\begin{tabular}{|l|l|l|}
		\hline
		\multicolumn{1}{|c|}{\multirow{2}{*}{\textbf{Code Transformations}}} & \multicolumn{2}{c|}{\textbf{Accu}} \\ \cline{2-3} 
		\multicolumn{1}{|c|}{} & \textbf{$ T_{cross} $} & \textbf{$ T_{cwe} $} \\ \hline
		Change Parameter & 126/495(25.5\%) & 17/49(34.7\%) \\ \hline
		Change Throw Exception & 98/227(43.1\%) & 5/15(33.3\%) \\ \hline
		Change Variable Definition & 63/265(23.8\%) & 11/33(33.3\%) \\ \hline
		Change Method Call & 41/194(21.1\%) & 4/11(36.4\%) \\ \hline
		Change Target & 19/123(15.4\%) & 2/13(15.4\%) \\ \hline
		Change String & 79/178(44.4\%) & 12/21(57.1\%) \\ \hline
		Change Method Declaration & 47/197(23.9\%) & 3/13(23.1\%) \\ \hline
		Change Class Declaration & 1/57(1.8\%) & 0/3(0\%) \\ \hline
		Change If Condition & 28/167(16.8\%) & 2/7(28.6\%) \\ \hline
		Change Switch block & 3/31(9.7\%) & 0/2(0\%) \\ \hline
		Change Loop Condition & 2/38(5.3\%) & 0/2(0\%) \\ \hline
		Change Return Statement & 31/180(17.2\%) & 4/14(28.6\%) \\ \hline
		Change Keywords "this/super" & 7/18(38.3\%) & 1/5(20.0\%) \\ \hline
		Change Try Block & 2/17(11.8\%) & 1/3(33.3\%) \\ \hline
		Change Catch Exception & 1/13(7.7\%) & 0/1(0\%) \\ \hline		
		Refactoring & 4/85(4.7\%) & 0/1(0\%) \\ \hline
		Other & 7/22(31.8\%) & 1/6(16.7\%) \\ \hline
	\end{tabular}
\end{table}

We can observe some conclusions from Table~\ref{tab:trans_result}.
In $ T_{cross} $, SeqTrans performs well in predicting throw exception, string, and keywords changes.
All of them substantially above average accuracy.
When predicting parameter change, method declaration, and variable definition. 
SeqTrans also performs better than the average accuracy.
In $ T_{cwe} $, SeqTrans performed consistently with $ T_{cross} $.
Only class declaration, switch block, loop condition, catch exception changes, and refactoring show lower accuracy than others.
We believe this gap can be explained in two points: code change sophistication and relevance.
There are certain templates for code changes like string and throw exceptions.
SeqTrans can more easily learn how to modify such changes from historical data.
But some of code transformations involve sophisticated code changes\footnote[1]{CVE-2015-5171, UAA, 9730cd6a3bbb481ee4e400b51952b537589c469d}, while others may only be due to insufficient samples, resulting in the model not learning well.
On the other hand, code changes such as refactorings and switch structure changes are difficult to accomplish with independent statement changes because the code is so interconnected.
This also leads to a decrease in model prediction accuracy.

\vspace{1em}
\noindent\fbox{
	\parbox{0.95\linewidth}{
		\textbf{Answer to RQ3. Finding 2:} SeqTrans performs well in handling throw exception change, string change and keywords change in both datasets. 	
		Simple code transformations is easier to be learned by the model, even in unseen situations.
		Sophisticated code and strongly correlated code transformations is not easily modified.
}}
\vspace{1em}

Overall, SeqTrans will perform well above average against specific kinds of CWE and specific kinds of code transformations.
As the model iterates in the hands of developers and the size of the data increases, we believe SeqTrans has much space for improvement.

\section{Discussion}
\label{sec:discussion}
\subsection{Internal Threats}
The performance of the NMT model can be significantly influenced by the hyperparameters we adopted. 
The transformer model is susceptible to hyperparameters. 
In order to mimic the Google setup, we set a bunch of options suggested by OpenNMT \cite{opennmt} to simulate their result.
However, there are gaps between source code language and natural language.
We also modified and tested part of the hyperparameters and chose the one that achieved the best performance.

We manually analyzed the prediction result and the source code, classified them into 17 types.
This number of categories is based on our experience during the experiment process, which may not be complete enough to cover all the code transformations.
More refined classification may lead to more discoveries.
However, during our analysis, we found that most code changes can be categorized into specific code transformations or a list of them.
Only a few code changes cannot be identified, classified, and even partly should be attributed to the mismatch of Gumtree \cite{falleri2014fine}.
In addition, there is the potential to introduce human error in the validation process.
We have taken our best efforts to avoid human errors.
All the validators in the experiments have more than three years of experience in Java development.

The small dataset and the complex transformer model may face the overfitting problem, which is occurred for three reasons: a small dataset, too many training steps and a complex model which is not fully trained.
In this work, we referenced He's work~\cite{he2019rethinking} and applied a pre-training model to alleviate it.
He's work proposes the following observation:
\begin{enumerate}
	\item Training from scratch is not a bad choice, either. 
	\item Pre-training allows the model to be converged earlier.
	\item When the amount of material is small, the pre-trained model is less likely to be over-fitted.
	\item Pre-training is helpless for tasks that are not very homogeneous.
\end{enumerate}
We think our specific domain dataset meets the above conditions. 
The transformer model is more complex than the seq2seq model, which contains more parameters to be fully trained.
Pre-training will speed up convergence on the target task.
Applying a pre-training model will be helpful to alleviate the overfitting problem. 
Our experimental results have also confirmed this opinion.

\subsection{External Validity}
During the experiment, we find that Gumtree \cite{falleri2014fine} will introduce mismatches, which will affect the quality of the training set.
Other researchers have mentioned that occasionally GumTree cannot appropriately detect motion and update actions between two ASTs~\cite{frick2018generating, matsumoto2019beyond}.
In fact, we found two problems with Gumtree, one is related to the IO issue. 
We found that the IO streams Gumtree used can cause blockages, and this has been confirmed and fixed by Gumtree’s author.
Another problem is in the bottom-up algorithm part of Gumtree.
This question did not receive a response from the author.
Neither did we do further experiment to evaluate the false-positive rate.
Verifying this problem is very difficult, and we have difficulty collecting a suitable ground truth.
We also modified Gumtree to support statement-level code matching and def-use chain collection.
We believe that through these, we have minimized the impact of Gumtree.

In addition, although we did not directly include fault localization in our evaluation of SeqTrans, we have also done some experiments related to fault location accuracy. 
We have investigated the popular fault localization tools and finally chose SpotBugs~\cite{spotbugs}.
It contains a plugin named Find Security Bugs~\cite{find-sec-bugs}, designed to detect 138 different vulnerability types with over 820 unique API signatures.
We have compared the bug reports provided by Spotbugs with our known vulnerability locations provided by the fix records.
Unfortunately, SpotBugs can only detect about 15\% of the vulnerability locations correctly. 
This result is beyond our expectations.
This low result shows that vulnerability localization is such a difficult work.
The latest automatic program repair tools can still only be used to assist developers.
There is still a long way to separate from the developers and independently do accurate automatic program fixes.
Exploring how to combine fault localization and automatic program repair together will be an important future work for us.

\subsection{Limitations}
The main limitation of SeqTrans is that it currently only supports the single-line prediction.
We always assume that these vulnerable statements are independent of each other when making predictions about the full CVEs.
We plan to abstract and tokenize the vulnerable function at the function-level, and the data format we currently use cannot handle this length quite well.

\subsection{Applications}
We believe SeqTrans can help programmers reduce repetitive work and give reasonable recommendations for fixing vulnerable statements.
As SeqTrans receives more and more modification records from developers, we believe there is still space for improvement in its performance.
We have also developed a VSCode plugin of SeqTrans to provide suggestions for developers to improve their codes, which will be opened soon.

On the other hand, training a generic model on large-scale data is very expensive, and it takes a long time to adjust the hyperparameters.
It would be meaningful work to provide a general model for subsequent researchers to refine directly based on this model.

The source code of SeqTrans is available at https://github.com/chijianlei/SeqTrans. 

This approach can also be applied to areas outside of vulnerability fixing, such as fine-grained code refactoring. 
We can use historical knowledge to refactor target code such as attribute extraction, merge parameter, inline variable, etc.
This is also part of our future exploration work.
Moreover, our study is based on the Java language now.
However, we believe that there is a common logic between programming languages, and the rules and features learned by the model can be easily applied to other languages.
\section{Related Works}
\label{sec:related}
In recent years, Deep Learning (DL) has become a powerful tool to solve problems of Software Engineering (SE), which can capture and discover features by the DL model rather than manual derivation.
In this work, we apply the Neural Machine Translation (NMT) model into the program repair field to learn from historical vulnerability repair records, summarize common pattern rules to apply to subsequent vulnerability fixes.
In the following, we will introduce studies focus on program repair and compare our work with related research.

\textbf{Automated Program Repair} 
Traditional program repair techniques can be categorized into two main categories: heuristic-based~\cite{goues2019automated}, constraint-based~\cite{goues2019automated}.
These techniques can sometimes be enhanced by machine learning, which we call learning-based repair~\cite{goues2019automated}.
It should be noted that the classification between these three approaches is vague, many techniques use more than one of them simultaneously.
We will list some traditional techniques to explain these three types of approaches.

\textit{Heuristic-based APR approaches} construct and traverse the search space for syntax program modifiers~\cite{goues2019automated}.
ARJA-e \cite{yuan2020toward} proposes a new evolutionary repair system for Java code that aims to address challenges for the search space.
SimFix~\cite{jiang2018shaping} utilizes both existing patches and similar code. 
It mines an abstract search space from existing patches and obtains a concrete search space by differencing with similar code snippets. 
Gatafix~\cite{bader2019getafix}  is based on a novel hierarchical clustering algorithm that summarizes fix patterns into a hierarchy ranging from general to specific patterns.
GenProg~\cite{weimer2009automatically} and RSRepair~\cite{qi2014strength} are two similar approaches.
Both of them try to repair faulty programs with the same mutation operations in a search space.
But GenProg uses random search, rather than genetic programming, to guide the patch generation process.
Meditor~\cite{xu2019meditor} provides a novel algorithm that flexibly locates and groups MR (migration-related) code changes in commits.
For edit application, Meditor matches a given program with inferred edits to decide which edit is applicable and produce a migrated version for developers.
AppEvolve~\cite{fazzini2019automated} can automatically perform app updates for API changes based on examples of how other developers evolved their apps for the same changes. 
This technique is able to update 85\% of the API changes considered, but it is quite time-consuming and not scalable enough.

Some approaches mine and learn fixing patterns from prior bug fixes. 
SimFix~\cite{jiang2018shaping}, FixMiner~\cite{koyuncu2020fixminer}, ssFix~\cite{xin2017leveraging}, CapGen~\cite{wen2018context} and HDRepair~\cite{le2016history}are based on frequently occurred code change operations that are extracted from the patches in code change histories.
The main difference between them is the object from which the data is extracted and how the data is processed.
AVATAR~\cite{liu2019avatar} exploits fix patterns of static analysis violations as ingredients for patch generation.
SOFix~\cite{liu2018mining} has a novel approach to digging up bug fix records from Stack Overflow responses.

These studies are still based on statistical ranking or strict context matching.
However, more and more studies are beginning to exploit machine learning to rank similar code transformations and automatically generate code recommendations.

\textit{Constraint-based APR approaches} usually focus on fixing a conditional expression, which is more prone to defects than other types of program elements.
Elixir~\cite{saha2017elixir} uses method call-related templates from par with local variables, fields or constants, to construct more expressive repair expressions, that go into synthesizing patches.
ACS~\cite{xiong2017precise} focuses on fine-grained ranking criteria for condition synthesis, which combines three heuristic ranking techniques that exploit the structure of the buggy program, the document of the buggy program, and the conditional expressions in existing projects. 

\textit{Learning-based APR approaches} is actually part of heuristic-based APR approaches that are enhanced by machine learning techniques.
We have separated them as an independent category.
DeepFix~\cite{gupta2017deepfix} is a program repair tool using a multi-layered sequence-to-sequence neural network with attention for fixing common programming errors.
In a collection of 6,971 incorrect C language programs written by students for 93 programming tasks, DeepFix can completely repair 1881 (27\%) of them, and can partially repair 1338 (19\%) of them.
HERCULES~\cite{saha2019harnessing} presents an APR technique that generalizes single-hunk repair techniques to include an important class of multi-hunk bugs, namely bugs that may require applying a substantially similar patch at a number of locations.
The limitation is that it addresses only a specific class of multi-hunk repairs and the evaluation is only carried out on the Defects4J dataset.
TRACER~\cite{ahmed2018compilation} is another work that is very similar to Deepfix for fixing compiler errors, and its accuracy rate exceeds that of Deepfix.
Tufano et al.~\cite{tufano2018empirical, tufano2019learning} has investigated the feasibility of using NMT for learning wild code. 
The disadvantage of his method is that only sentences with less than 100 tokens are analyzed.
In addition, this work is only limited to the type of bug that contains only one sequence within a single method.

SequenceR~\cite{chen2019sequencer} presents a novel end-to-end approach to program repair based on sequence-to-sequence learning.
It utilizes the copy mechanism to overcome the unlimited vocabulary problem. 
To the best of our knowledge, it achieves the best result reported on such a task.
However, the abstract data structure of this method retains too much useless context.
It does not use the normalization method either.

\textbf{Vulnerability Repair}
Fixing vulnerability is critical to protect users from security compromises and prevent vendors from losing user confidence. 
Traditional tools such as Angelix~\cite{mechtaev2016angelix}, Semfix~\cite{nguyen2013semfix} and ClearView~\cite{perkins2009automatically} heavily rely on a set of positive/negative example inputs to find a patch that makes the program behaves correctly on those examples.
SENX~\cite{huang2019using} propose a different approach called ``property-based" which relies on program-independent, vulnerability-specific, human-specified safety properties.

Another trending direction is the application of neural network models for vulnerability repair.
Harer et al.~\cite{harer2018learning} apply Generative Adversarial Network (GAN) to the problem of automated repair of software vulnerabilities. 
They address the environment with no labeled vulnerable examples and achieve performance close to seq2seq approaches
that require labeled pairs.
Chen et al.~\cite{chen2019using} apply the simple seq2seq model for vulnerability repair but the performance is not quite promising.
Ratchet~\cite{hata2018learning} also utilizes the NMT model to fix vulnerabilities, but it only stores single statements without any context around them.
All of these functions do not consider multiple-statement, either.

\textbf{Transformer and Tree Structure} Another popular direction is utilizing the deep learning model or treating source code as a syntax tree to maintain richer information. 
Tran$ S^3 $ \cite{wang2020trans} proposes a transformer-based framework to integrate code summarization with code search.
Tree-based neural network such as TreeLSTM~\cite{ahmed2019improving, tai2015improved}, ASTNN~\cite{zhang2019novel} or TreeNet~\cite{cheng2018treenet} are also being applied on program analysis.
Shiv et al.~\cite{shiv2019novel} propose a method to extend transformers to tree-structured data.
This approach abstracts the sinusoidal positional encodings of the transformer, using a novel positional encoding scheme to represent node positions within trees.
It achieves a 22\% absolute increase in accuracy on a JavaScript to
CoffeeScript~\cite{burnham2015coffeescript} translation dataset.
TreeCaps~\cite{jayasundara2019treecaps} proposes a tree-based capsule network for processing program code in an automated way that encodes syntactical code structures and captures code dependencies more accurately.
CODIT~\cite{chakraborty2020codit} and DLFix~\cite{li2020dlfix} has begun to apply tree structure into program repair and achieve some progress.

The most similar work to us is VRepair~\cite{chen2021neural}.
Both of the two studies used fine-tuning to solve the small sample problem.
The size of their training set is also in the same order of magnitude as ours.
The main differences between VRepair and SeqTrans are the targeted languages and data structures.
VRepair focuses on the C language but the target of SeqTrans is on the Java language.
Also, in order to decreases the size of the output sequence, VRepair represents edit scripts at token level and the network only outputs the changed source code tokens not the whole function.
However, the problem is that multiple inference results will be generated when backfilling the modified token.
In our approach, we will maintain the suspicious statements and all statements that contain data dependencies with the suspicious statements.
In other words, we will preserve more context around the suspicious statements but also make sequences longer.
In addition, his work does not provide a runnable example or code.

Most of these techniques focus on single statement prediction.
Translating multiple statements together is more challenging than translating one language to another language. 
Techniques for characterizing code using tree and graph structures and converting the resulting prediction trees into readable code are still in the exploratory stage.
Overall, we believe that using a tree-based neural network or even combining it with a transformer structure will become our future work.

\section{Conclusion}
\label{sec:conclusion}
In this paper, we design the automatic vulnerability fix tool SeqTrans based on the NMT technique to learn from historical vulnerability fixes.
It can provide suggestions and automatically fix the source code for developers.
Fine-tuning strategy is used to overcome the small sample size problem.
We conduct our study on real-world vulnerability fix records and compare our SeqTrans with three kinds of other NMT techniques.
We investigated three research questions based on these collected data.
Experiment results show that our technique outperforms the state-of-the-art NMT model and achieves an accuracy rate of 23.3\% in statement-level prediction and 25.3\% in CVE-level prediction.
The SeqTrans-based approach indeed helps solve the scalability and small data set problems of existing methods on the task of vulnerability fixing.
We also look deeply into the model and manually analyze the prediction result and the source code.
Our observation finds that SeqTrans performs exceptionally well in specific kinds of CWEs like CWE-287 (Improper Authentication) and CWE-863 (Incorrect Authorization). The prediction range will become wider and wider as the historical repair records increases.

\section{Acknowledgement}
This work was supported by National Key Research and Development Program of China (2018YFB1004500), National Natural Science Foundation of China (62002280, 61632015, 61772408, U1766215, 61833015,61902306), Innovative Research Group of the National Natural Science Foundation of China (61721002), Innovation Research Team of Ministry of Education (IRT\_17R86), Project of China Knowledge Centre for Engineering Science and Technology. Project of Chinese Academy of Engineering ``The Online and Offline Mixed Educational ServiceSystem for 'The Belt and Road' Training in MOOC China''

\bibliographystyle{IEEEtran}
\bibliography{references}


\begin{IEEEbiography}[{\includegraphics[width=1in,height=1.25in,clip,keepaspectratio]{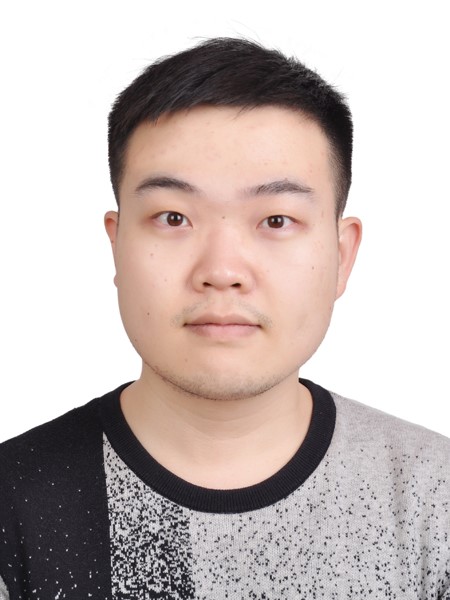}}]{Jianlei Chi}
received the B.S. degree in computer science and technology from Harbin Engineering University, China, 2014, and the Ph.D. degree in computer science and technology in 2022 from Xi’an Jiaotong University, China. He is a post-doctoral researcher at the Institute of Cyberspace Security, Zhejiang University of Technology, China. His research interests include trustworthy software, software engineering, program analysis and machine learning.
\end{IEEEbiography}

\begin{IEEEbiography}[{\includegraphics[width=1in,height=1.25in,clip,keepaspectratio]{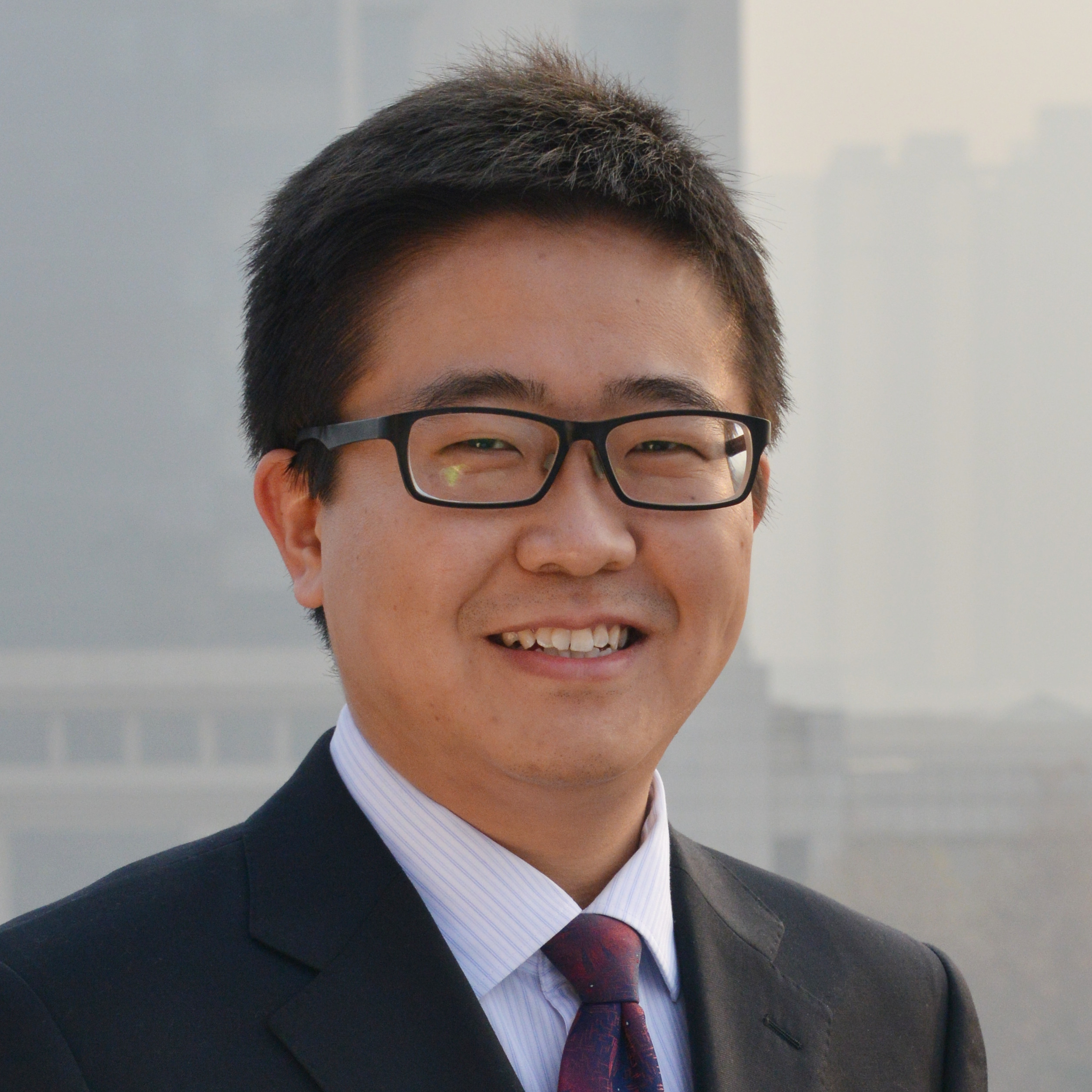}}]{Yu Qu}
received the B.S. and Ph.D. degrees from Xi’an Jiaotong University, Xi’an, China in 2006 and 2015 respectively. He is a post-doctoral researcher at the Department of Computer Science and Engineering, UC Riverside. His research interests include trustworthy software and applying complex network and data mining theories to analyzing software systems.
\end{IEEEbiography}

\begin{IEEEbiography}[{\includegraphics[width=1in,height=1.25in,clip,keepaspectratio]{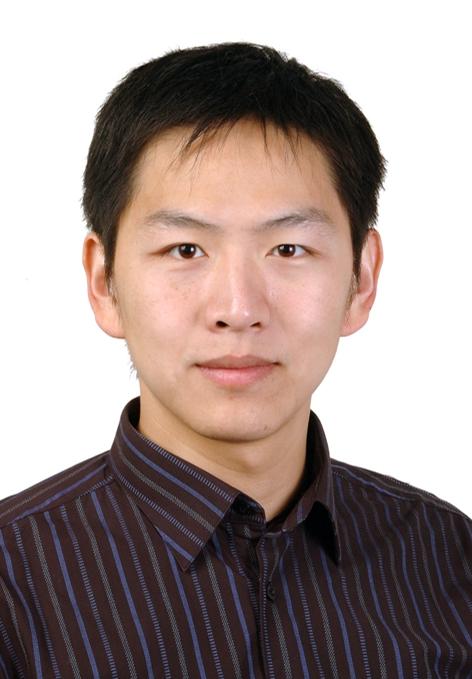}}]{Ting Liu}
received his B.S. degree in information engineering and Ph.D. degree in system engineering from School of Electronic and Information, Xi’an Jiaotong University, Xi’an, China, in 2003 and 2010, respectively. Currently, he is a professor of the Systems Engineering Institute, Xi’an Jiaotong University. His research interests include smart grid, network security and trustworthy software.
\end{IEEEbiography}

\begin{IEEEbiography}[{\includegraphics[width=1in,height=1.25in,clip,keepaspectratio]{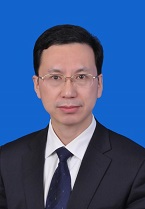}}]{Qinghua Zheng}
Qinghua Zheng received the B.S. degree incomputer software in 1990, the M.S. degree in computer organization and architecture in 1993, and the Ph.D. degree in system engineering in 1997 from Xi’an Jiaotong University, China. He was a postdoctoral researcher at Harvard University in 2002. He is currently a professor in Xi’an Jiaotong University, and the dean of the Department of Computer Science.
His research areas include computer network security, intelligent e-learning theory and algorithm, multimedia e-learning, and trustworthy software.
\end{IEEEbiography}

\begin{IEEEbiography}[{\includegraphics[width=1in,height=1.25in,clip,keepaspectratio]{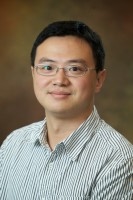}}]{Heng Yin}
is a professor in the department of Computer Science and Engineering at UC Riverside. Before joining UC Riverside, he was with Syracuse University from September 2009 to June 2016, as assistant professor and then associate professor. He obtained his Ph.D in Computer Science from the College of William and Mary in 2009, while He spent 4 years at Carnegie Mellon University and later at UC Berkeley.
His research interests lie in computer security and developing all kinds of techniques (such as program analysis, virtualization, and machine learning/deep learning) to solve computer and software security problems, including but not limited to malware detection and analysis, vulnerability discovery, program hardening, digital forensics.
\end{IEEEbiography}

\end{document}


\begin{table*}[htbp]
	\renewcommand{\arraystretch}{1.1}
	\caption{Description of all CWEs in Ponta's dataset}
	\label{tab:appendix}
	\centering
	\scriptsize
	\begin{tabular}{|l|l|}
		\hline
		\textbf{CWE ID} & \textbf{Description} \\ \hline
		CWE-6 & Improper Restriction of XML External Entity Reference \\ \hline
		CWE-16 & Configuration \\ \hline
		CWE-19 & Data Processing Errors \\ \hline
		CWE-20 & Improper Input Validation \\ \hline
		CWE-22 & Improper Limitation of a Pathname to a Restricted Directory ('Path Traversal') \\ \hline
		CWE-74 & Improper Neutralization of Special Elements in Output Used by a Downstream Component ('Injection') \\ \hline
		CWE-77 & Improper Neutralization of Special Elements used in a Command ('Command Injection') \\ \hline
		CWE-78 & Improper Neutralization of Special Elements used in an OS Command ('OS Command Injection') \\ \hline
		CWE-79 & Improper Neutralization of Input During Web Page Generation ('Cross-site Scripting') \\ \hline
		CWE-89 & Improper Neutralization of Special Elements used in an SQL Command ('SQL Injection') \\ \hline
		CWE-91 & XML Injection (aka Blind XPath Injection) \\ \hline
		CWE-93 & Improper Neutralization of CRLF Sequences ('CRLF Injection') \\ \hline
		CWE-94 & Improper Control of Generation of Code ('Code Injection') \\ \hline
		CWE-113 & Improper Neutralization of CRLF Sequences in HTTP Headers ('HTTP Response Splitting') \\ \hline
		CWE-119 & Improper Restriction of Operations within the Bounds of a Memory Buffer \\ \hline
		CWE-123 & Write-what-where Condition \\ \hline
		CWE-184 & Incomplete Blacklist \\ \hline
		CWE-189 & Numeric Errors \\ \hline
		CWE-190 & Integer Overflow or Wraparound \\ \hline
		CWE-200 & Information Exposure \\ \hline
		CWE-203 & Information Exposure Through Discrepancy \\ \hline
		CWE-209 & Information Exposure Through an Error Message \\ \hline
		CWE-212 & Improper Cross-boundary Removal of Sensitive Data \\ \hline
		CWE-254 & 7PK - Security Features \\ \hline
		CWE-255 & Credentials Management \\ \hline
		CWE-264 & Permissions Privileges and Access Controls \\ \hline
		CWE-269 & Improper Privilege Management \\ \hline
		CWE-284 & Improper Access Control \\ \hline
		CWE-285 & Improper Authorization \\ \hline
		CWE-287 & Improper Authentication \\ \hline
		CWE-295 & Improper Certificate Validation \\ \hline
		CWE-297 & Improper Validation of Certificate with Host Mismatch \\ \hline
		CWE-306 & Missing Authentication for Critical Function \\ \hline
		CWE-310 & Cryptographic Issues \\ \hline
		CWE-319 & Cleartext Transmission of Sensitive Information \\ \hline
		CWE-320 & Key Management Errors \\ \hline
		CWE-327 & Use of a Broken or Risky Cryptographic Algorithm \\ \hline
		CWE-345 & Insufficient Verification of Data Authenticity \\ \hline
		CWE-347 & Improper Verification of Cryptographic Signature \\ \hline
		CWE-352 & Cross-Site Request Forgery (CSRF) \\ \hline
		CWE-354 & Improper Validation of Integrity Check Value \\ \hline
		CWE-358 & Improperly Implemented Security Check for Standard \\ \hline
		CWE-361 & 7PK - Time and State \\ \hline
		CWE-362 & Concurrent Execution using Shared Resource with Improper Synchronization ('Race Condition') \\ \hline
		CWE-384 & Session Fixation \\ \hline
		CWE-388 & 7PK - Errors \\ \hline
		CWE-399 & Resource Management Errors \\ \hline
		CWE-400 & Uncontrolled Resource Consumption \\ \hline
		CWE-404 & Improper Resource Shutdown or Release \\ \hline
		CWE-417 & Channel and Path Errors \\ \hline
		CWE-434 & Unrestricted Upload of File with Dangerous Type \\ \hline
		CWE-444 & Inconsistent Interpretation of HTTP Requests ('HTTP Request Smuggling') \\ \hline
		CWE-470 & Use of Externally-Controlled Input to Select Classes or Code ('Unsafe Reflection') \\ \hline
		CWE-502 & Deserialization of Untrusted Data \\ \hline
		CWE-521 & Weak Password Requirements \\ \hline
		CWE-522 & Insufficiently Protected Credentials \\ \hline
		CWE-532 & Inclusion of Sensitive Information in Log Files \\ \hline
		CWE-592 & DEPRECATED: Authentication Bypass Issues \\ \hline
		CWE-601 & URL Redirection to Untrusted Site ('Open Redirect') \\ \hline
		CWE-611 & Improper Restriction of XML External Entity Reference \\ \hline
		CWE-613 & Insufficient Session Expiration \\ \hline
		CWE-640 & Weak Password Recovery Mechanism for Forgotten Password \\ \hline
		CWE-668 & Exposure of Resource to Wrong Sphere \\ \hline
		CWE-732 & Incorrect Permission Assignment for Critical Resource \\ \hline
		CWE-755 & Improper Handling of Exceptional Conditions \\ \hline
		CWE-770 & Allocation of Resources Without Limits or Throttling \\ \hline
		CWE-776 & Improper Restriction of Recursive Entity References in DTDs ('XML Entity Expansion') \\ \hline
		CWE-789 & Uncontrolled Memory Allocation \\ \hline
		CWE-829 & Inclusion of Functionality from Untrusted Control Sphere \\ \hline
		CWE-835 & Loop with Unreachable Exit Condition ('Infinite Loop') \\ \hline
		CWE-863 & Incorrect Authorization \\ \hline
		CWE-918 & Server-Side Request Forgery (SSRF) \\ \hline
		CWE-1188 & Insecure Default Initialization of Resource \\ \hline
		CWE-noinfo & Insufficient Information \\ \hline
		CWE-Other & Other \\ \hline
	\end{tabular}
\end{table*}